\documentclass[12pt]{article} 
\pdfoutput=1
\usepackage{amssymb,graphicx}
\usepackage{epstopdf}
\usepackage{amsmath,amsfonts}
\usepackage{epsfig} 
\usepackage{graphicx,graphics}
\usepackage{xcolor}
\usepackage{hyperref}
\usepackage{soul}
\usepackage[font=small,labelfont=bf]{caption}
\newcommand{\CLns}{{\tt ${\mathcal C}$osmo${\mathcal L}$attice}}
\begin{document} 

\title{\bf Revisiting evolution of domain walls and their gravitational radiation with CosmoLattice}
\author{I.~Dankovsky$^{a,b}$, E.~Babichev$^c$, D.~Gorbunov$^{b, d}$, S.~Ramazanov$^e$, A.~Vikman$^f$\\
\small{\em $^a$Faculty of Physics, Lomonosov Moscow State University,}\\
\small{\em  Leninskiye Gory 1-2, 119991 Moscow, Russia}\\
\small{\em $^b$Institute for Nuclear Research of the Russian Academy of Sciences, 117312 Moscow, Russia}\\
\small{\em $^c$Universit\'e Paris-Saclay, CNRS/IN2P3, IJCLab, 91405 Orsay}\\
\small{\em $^d$Moscow Institute of Physics and Technology, 141700 Dolgoprudny, Russia}\\
\small{\em  $^e$Institute for Theoretical and Mathematical Physics, MSU, Moscow 119991, Russia}\\
\small{\em $^f$CEICO, FZU-Institute of Physics of the Czech Academy of Sciences,}\\
\small{\em Na Slovance 1999/2, 182 00 Prague 8, Czech Republic}
}
 
 \date{}

{\let\newpage\relax\maketitle}

\begin{abstract}

Employing the publicly available \CLns~code, we conduct numerical simulations of a domain wall network and the resulting gravitational waves (GWs) in a radiation-dominated Universe in the $Z_2$-symmetric scalar field model. In particular, the domain wall evolution is investigated in detail both before and after reaching the scaling regime, using the combination of numerical and theoretical methods. We demonstrate that the total area of closed walls is negligible compared to that of a single long wall stretching throughout the simulation box. Therefore, the closed walls are unlikely to have a significant impact on the overall network evolution. This is in contrast with the case of cosmic strings, where formation of loops is crucial for maintaining the system in the scaling regime. To obtain the GW spectrum, we develop a technique that separates physical effects from numerical artefacts arising due to finite box size and non-zero lattice spacing. Our results on the GW spectrum agree well with Refs.~\cite{Kitajima:2023cek, Ferreira:2023jbu}, which use different codes. Notably, we observe a peak at the Hubble scale, an exponential falloff at scales shorter than the wall width, and a plateau/bump at intermediate scales. We also study sensitivity of obtained results on the choice of initial conditions. We find that different types of initial conditions lead to qualitatively similar domain wall evolution in the scaling regime, but with important variations translating into different intensities of GWs.

\end{abstract}

\sloppy

\section{Introduction}

\label{sec:intro}

The first evidence of stochastic gravitational wave (GW) background has been recently found by several collaborations including NANOGrav~\cite{NANOGrav:2023gor}, EPTA\footnote{PTA stands for Pulsar Timing Array.} with InPTA~\cite{EPTA:2023fyk}, PPTA~\cite{Reardon:2023gzh}, and Chinese PTA~\cite{Xu:2023wog}. 
While supermassive black hole binaries (SMBHBs) are often thought to be the most likely source of the signal detected~\cite{NANOGrav:2023hfp, EPTA:2023xxk}, the latter can be also explained by various primordial mechanisms~\cite{NANOGrav:2023hvm, EPTA:2023xxk}. Moreover, the explanation invoking SMBHBs may require highly nontrivial astrophysical mechanisms, see, e.g., Refs.~\cite{Sesana:2013wja, Ellis:2023oxs,Sato-Polito:2023gym}. 
In light of this and future searches for stochastic GW backgrounds with LISA~\cite{LISA:2017pwj} and the Einstein Telescope~\cite{Hild:2010id}, it is timely to investigate the fine structure of the GW spectrum generated by a specific primordial source. This detailed analysis could enable us to distinguish it from the GW spectrum of astrophysical objects. In this work we focus on GWs from evolving cosmic domain walls~\cite{Zeldovich:1974uw}. 

Domain walls arise in field theories with scalars which transform under discrete symmetries spontaneously broken by non-zero vacuum expectation values. For concreteness, we focus on the simplest model of $Z_2$-symmetric scalar field $\phi$:
\begin{equation}
\label{Lagrangian}
S=\int d^4 x \sqrt{-g} \left[\frac{1}{2} g^{\mu \nu}\partial_{\mu} \phi \partial_{\nu} \phi-
\frac{\lambda}{4} \left(\phi^2 -\eta^2\right)^2 \right] \; .
\end{equation}
Here $\eta$ is the field $\phi$ expectation value spontaneously breaking $Z_2$-symmetry. 
Domain walls are often viewed as problematic in cosmology because, while the Universe expands, their energy density grows relative to the energy density of the surrounding matter. Thus, unless their tension is chosen to be very small~\cite{Zeldovich:1974uw, Lazanu:2015fua}, domain walls would promptly become the dominant driving source for cosmological evolution, invalidating the standard expansion history. This problem is cured, e.g., by adding terms, which explicitly break $Z_2$-symmetry~\cite{Zeldovich:1974uw, Gelmini:1988sf}, introducing biased initial conditions~\cite{Coulson:1995nv}, or by promoting 
$\eta$ to a dynamical quantity~\cite{Vilenkin:1981zs, Ramazanov:2021eya, Babichev:2021uvl, Babichev:2023pbf}. In this work, however, 
we do not discuss the domain wall problem 
assuming that it is resolved in one or another way, so that the domain wall network gets destroyed at some moment of time in the early Universe without any contradiction with  cosmological observations.

 Evolution of domain walls has been first studied numerically in Ref.~\cite{Press:1989yh}. To make the problem tractable, a non-physical assumption was made that the wall width grows linearly with the scale factor. 
It was concluded in Ref.~\cite{Press:1989yh} that the domain wall 
network enters a scaling regime with one or a few long domain walls per horizon volume. The scaling regime was confirmed in later simulations~\cite{Hiramatsu:2013qaa, Saikawa:2017hiv}, which, contrary to~\cite{Press:1989yh}, kept the wall width constant. Furthermore, Ref.~\cite{Hiramatsu:2013qaa} estimated the energy density and spectrum (frequency distribution) of GWs emitted by domain walls. The latter also serve as a source of particle emission studied in the context of axion string-wall networks~\cite{Hiramatsu:2012sc, Hiramatsu:2012gg, Kawasaki:2014sqa, Chang:2023rll}. 
Most recently, numerical simulations of domain walls have been focused on proper incorporation of 
the potential bias~\cite{Krajewski:2021jje, Kitajima:2023cek, Ferreira:2023jbu, Kitajima:2023kzu}, which triggers formation of primordial black holes through the collapse of false vacuum bubbles~\cite{Ferreira:2024eru, Dunsky:2024zdo, Gouttenoire:2023gbn}. Note that besides 
numerical approaches, (semi)-analytical modelling of domain walls is also 
being actively developed~\cite{Hindmarsh:1996xv, Avelino:2005kn, Martins:2016ois, Martins:2016lzc, Avelino:2020ubr, Pujolas:2022qvs, Gruber:2024pqh}.

\begin{figure}
    \includegraphics[width=\textwidth]{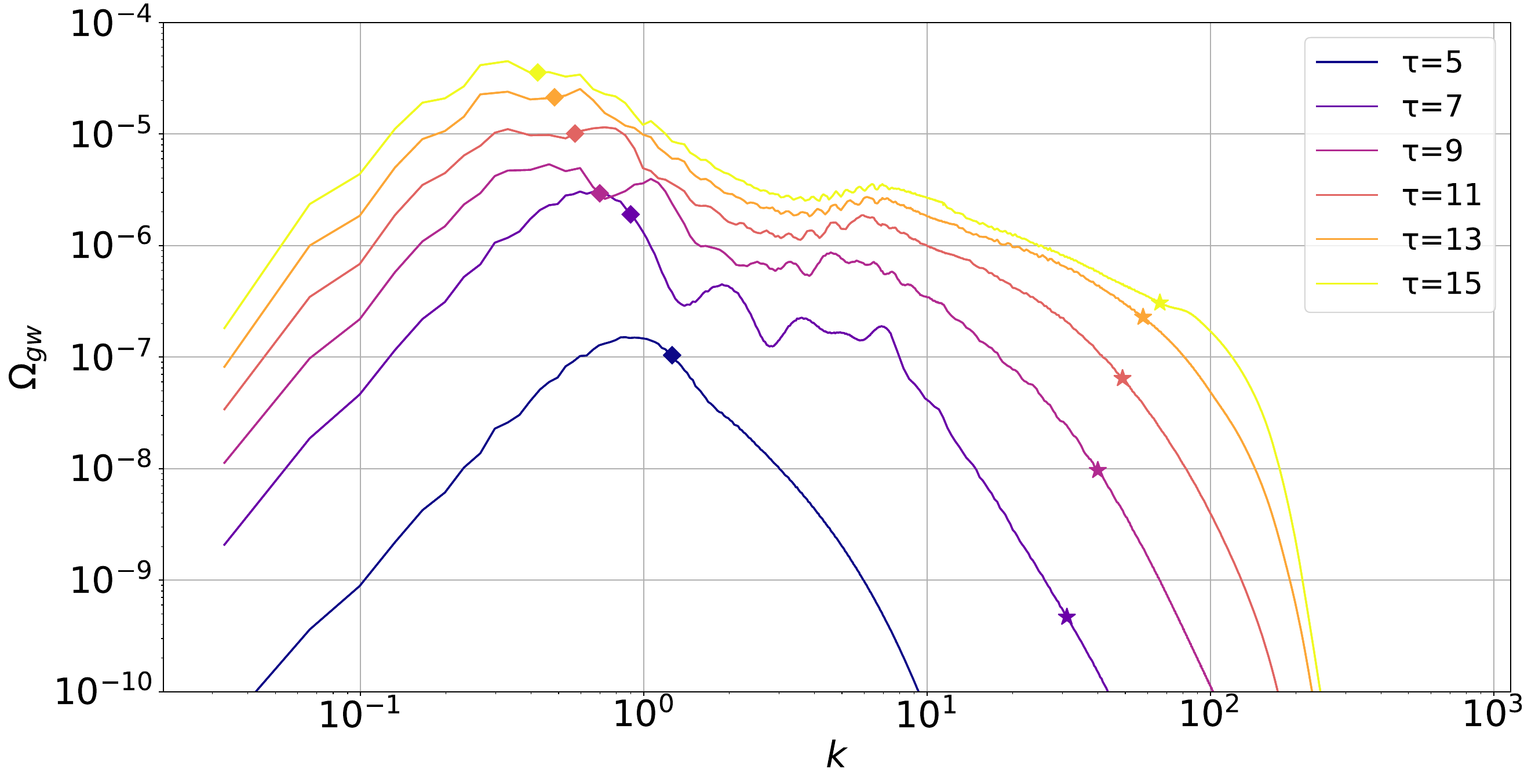}\\
        \includegraphics[width=\textwidth]{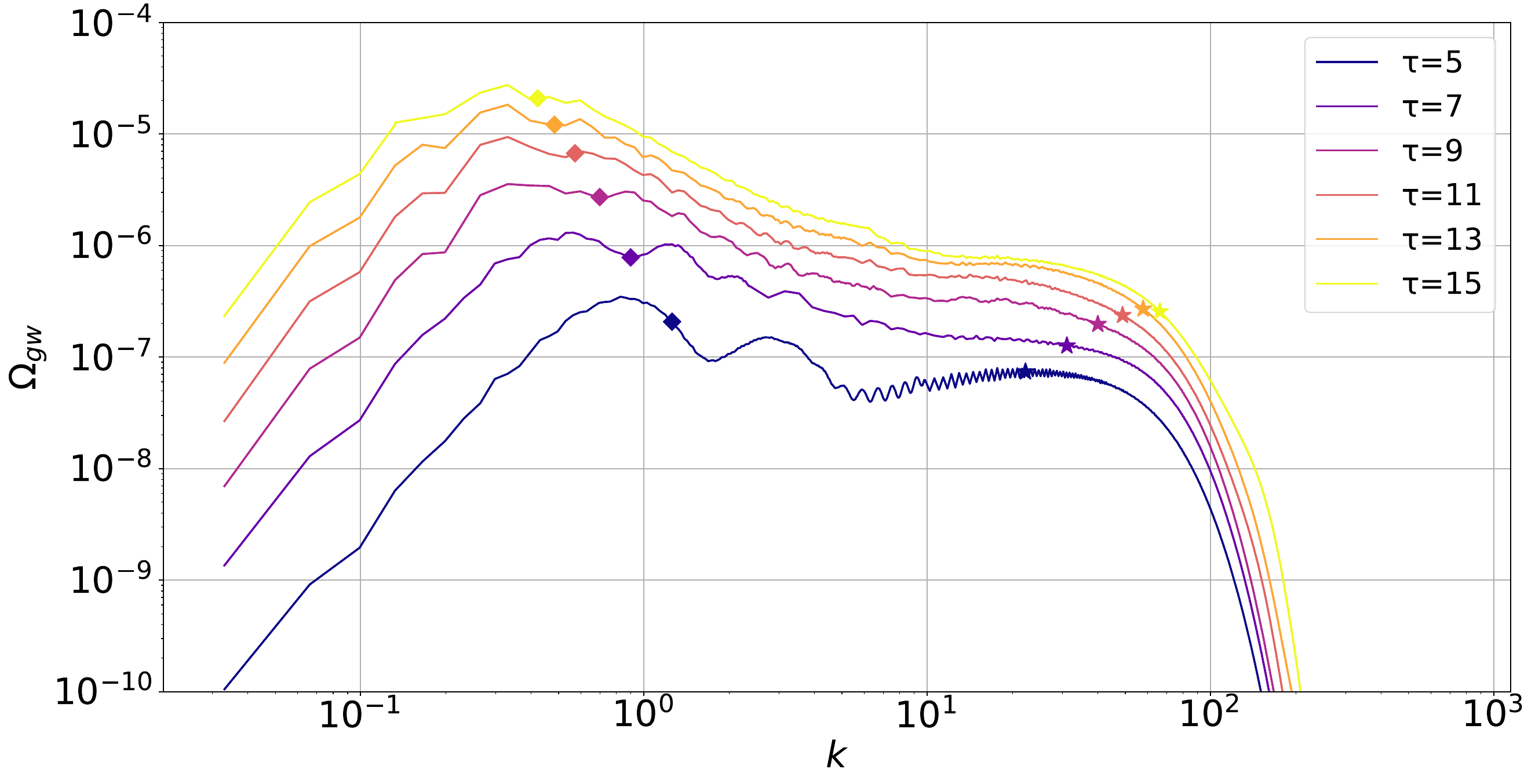}
    \caption{Spectrum of GWs emitted by the domain wall network at radiation domination starting with vacuum (top panel) and thermal (bottom panel) initial conditions defined in Eqs.~\eqref{vacuum_cond} and \eqref{thermal_cond}, respectively. Conformal momenta and conformal times are in units of $\sqrt{\lambda} \eta$ and $\frac{1}{\sqrt{\lambda}\eta}$, respectively. The sharp upper cutoff at $k_{cut}=1$ is applied in the case of vacuum initial conditions. The expectation value $\eta$ is set at $\eta=6 \cdot 10^{16}~\mbox{GeV}$. Rescaling to arbitrary $\eta$ is achieved by multiplying the spectra by $(\eta/6 \cdot 10^{16}~\mbox{GeV})^4$. Simulations have been performed on a lattice with a grid number $N=2048$. The positions of diamonds correspond to the comoving Hubble scale $k=2\pi H a$ at the time associated with the corresponding curves, while stars show the inverse domain wall width $1/\delta_{w}$, i.e., $k=2\pi a/\delta_{w}$.} \label{Finalspectra}
\end{figure}

The main goal of this work is to perform an independent study of domain wall evolution and the resulting GW production at radiation domination, with the help of the publicly available code \CLns~\cite{Figueroa:2020rrl, Figueroa:2021yhd, Figueroa:2023xmq}. Since the latter can be used by a broad audience, it is an important task to analyse, whether the results obtained with \CLns~and different codes~\cite{Hiramatsu:2013qaa, Kitajima:2023cek, Ferreira:2023jbu}, as well as analytical results of Sec.~\ref{sec:scaling}, agree between each other. Most simulations are performed assuming either vacuum initial conditions (with a cutoff at large momenta) or thermal ones defined in Eqs.~\eqref{vacuum_cond} and \eqref{thermal_cond}, respectively.

The structure of the domain wall network revealed with \CLns~is as follows. We observe in Sec.~\ref{sec:resultsdw} that the network mainly consists of a single wall stretching throughout the box, 
which simulates the early Universe comprising multiple Hubble patches. At the same time, closed walls constitute 
only a very small fraction of the total network, see also Ref.~\cite{Garagounis:2002kt}. We demonstrate this explicitly for the first time in the literature using histograms, see Fig.~\ref{Histograms} in Sec.~\ref{sec:resultsdw}. Neglecting closed walls, there is on average one long domain wall per Hubble volume in the scaling regime, --- in accordance with the previous studies~\cite{Press:1989yh, Hiramatsu:2013qaa}. We also provide a quantitative analysis of the domain wall scaling. We find that the scaling is reached when the wall width becomes $\sim 1/20$ fraction of the horizon radius\footnote{We assume here that the field $\phi$ starts rolling to minima when its tachyonic mass $\sqrt{\lambda} \eta$, which is of the order of the inverse wall width, becomes larger than the Hubble rate.}. This fact is remarkably independent of the choice of initial conditions, even though approaching the scaling solution may occur in drastically different ways. On the other hand, the domain wall area is slightly sensitive to the choice of initial conditions. The similar observation has been made in Ref.~\cite{Kitajima:2023kzu}, when comparing thermal and inflationary initial conditions. 
We prove that this difference 
is independent of lattice spacing, box size, and UV properties of the initial spectra, and hence it is likely to have a physical origin.

Previous simulations of GWs from domain walls using \CLns~\cite{Li:2023yzq, Li:2023gil} indicated strong departures from the results of Ref.~\cite{Hiramatsu:2013qaa}. In particular, Refs.~\cite{Li:2023yzq, Li:2023gil} noticed appearance 
of a pronounced peak in the ultraviolet (UV) part of the spectrum. 
We also observe the UV peak as the result of our simulations, but we prove that it is an artefact of the non-zero lattice spacing. We propose a technique to attain trustworthy predictions of the GW spectrum by comparing GW spectra obtained with different space resolutions. For this purpose we use $N\times N\times N$ lattices with different grid numbers $N=512$, $N=1024$, and $N=2048$.
 
 Our main results for the GW differential spectra, defined through the energy density of GWs $\rho_{gw}(\tau)$ and total energy density $\rho(\tau)$ as 
 \[
 \Omega_{gw}(\tau)\equiv\frac{d\rho_{gw}(\tau)}{\rho(\tau) \,d\log k} \; ,
 \]
 are summarised in Fig.~\ref{Finalspectra} at various conformal times $\tau$ and conformal momenta $k$ for vacuum (with a cutoff at large momenta) and thermal initial conditions. At this stage we do not switch on the mechanism of the domain wall destruction (which is needed for the healthy cosmology) and consequently ignore its possible impact on the GW production. The overall growth of the spectrum reflects constancy of GWs produced by domain walls in the scaling regime, which we prove analytically in Sec.~\ref{subsec:gwscaling}. The infrared (IR) tail of the GW spectrum as well as the behaviour in the vicinity of the main (scaling) peak are in a good agreement with those of Refs.~\cite{Hiramatsu:2013qaa, Kitajima:2023cek, Ferreira:2023jbu}. Nevertheless, the spectral index on the right of the peak is slightly different compared to Refs.~\cite{Hiramatsu:2013qaa, Kitajima:2023cek, Ferreira:2023jbu}, as we demonstrate in Sec.~\ref{sec:gw}. At intermediate scales, we observe a plateau for the thermal initial conditions and a bump for the vacuum initial conditions. Such features have been absent in the earlier simulations~\cite{Hiramatsu:2013qaa}, but can be seen in later analyses of Refs.~\cite{Kitajima:2023cek, Ferreira:2023jbu}. Note that presence of the bump in the case of vacuum initial conditions appears to be the consequence of the sharp cutoff. Indeed, artificially applying the sharp cutoff to thermal initial conditions also leads to the bump feature, cf. Fig.~\ref{GWthermalcutoff}.
Our simulations reveal other differences of GW spectra depending on the choice of initial conditions. In particular, at any time after\footnote{Note that $\tau=5$ is just the onset of the scaling regime, see Fig.~\ref{ksi} and Eq.~\eqref{scaling_onset}.} $\tau=5$ the heights of GW peaks on the top plot of Fig.~\ref{Finalspectra} exceed those on the bottom plot. That tendency can be traced back to the aforementioned difference of domain wall areas for the two types of initial conditions, see also Sec.~\ref{sec:resultsdw}.

The outline of the paper is as follows. In Sec.~\ref{sec:scaling}, we provide a theoretical background for domain wall evolution in the scaling regime and the associated GW emission. In Sec.~\ref{sec:prep}, we set up the strategy for the optimal analysis of the domain wall network on a lattice. In particular, we establish the optimal parameters to master the natural limitations such as a finite box size and a non-zero lattice spacing. 
We present results of simulations of the 
domain wall evolution and GW production in Secs.~\ref{sec:resultsdw} and~\ref{sec:gw}, respectively. We conclude in Sec.~\ref{sec:discussions} with a summary and discussion of the future prospects. 

\section{Scaling regime: analytic approach}

In this section we discuss analytic approach to domain wall network evolution and the resulting GWs. On scales much longer than the width, domain walls evolve towards a self-similar and scale-free 
("scaling") dynamical attractor regime described in Sec.~\ref{subsec:scaling}. 
In Sec.~\ref{subsec:gwscaling}, we calculate the energy density of GWs from domain walls, and show that it is constant in the scaling regime. Unlike Ref.~\cite{Hiramatsu:2013qaa}, we demonstrate this analytically without referring to the Einstein quadrupole formula. We also briefly review, how the spectral shape of the IR part of GW spectrum can be deduced solely from generic considerations of causality.

\label{sec:scaling}

\subsection{Domain walls in the scaling regime}

\label{subsec:scaling}

Let us consider some stochastic quantity $s({\bf x})$ describing the domain wall network and having zero expectation value, i.e., $\langle s({\bf x}) \rangle =0$, where we average over many realisations of the network with the same microscopic physics and cosmological conditions.  This could be, e.g., the 
expectation value of the scalar field $\phi$ comprising domain walls. Another example will be given in Sec.~\ref{subsec:gwscaling}. We characterise statistical properties of the field $s({\bf x})$ by the normalised 2-point function:
\begin{equation}
\label{dva}
\frac{\langle s({\bf x}, \tau) s({\bf y}, \tau') \rangle }{\sqrt{\langle s^2 ({\bf x}, \tau) \rangle \langle s^2({\bf y}, \tau') \rangle }} \equiv F({\bf x}, {\bf y}, \tau, \tau') \; .
\end{equation}
Accounting for the statistical homogeneity and isotropy, one can write 
\begin{equation}
F({\bf x}, {\bf y}, \tau, \tau') =F(|{\bf x}-{\bf y}|, \tau, \tau') \; .
\end{equation}
In terms of the Fourier transformed quantity,   
\begin{equation}
s({\bf k})=\frac{1}{(2\pi)^3}\int d {\bf x} ~e^{-i {\bf kx}} \cdot s({\bf x}) \; ,
\end{equation}
the correlation function respecting these properties has the form:
\begin{equation}
\label{long}
\begin{split}
F({\bf k}, {\bf q}, \tau, \tau') & \equiv \frac{1}{(2\pi)^6}\int d {\bf x}~e^{-i {\bf kx}}\int d {\bf y}~e^{-i {\bf qy}}   \cdot F({\bf x}, {\bf y}, \tau, \tau') \\
& = \frac{\langle s({\bf k}, \tau) s({\bf q}, \tau') \rangle}{\sqrt{\langle s^2 (\tau) \rangle \cdot \langle s^2 (\tau') \rangle}}
= \delta ({\bf k}+{\bf q}) P(k, \tau, \tau') \; ,
\end{split}
\end{equation}
where the power spectrum $P(k, \tau, \tau')$ is defined as 
\begin{equation}
\label{powerfull}
P(k, \tau, \tau') = \frac{1}{(2\pi)^3}\int d{\bf z} e^{-i{\bf kz}} F(|{\bf z}|, \tau, \tau') \; .
\end{equation}
We also defined ${\bf z} \equiv {\bf x}-{\bf y}$. In Eq.~\eqref{long}, we used that $\langle s^2 ({\bf x},\tau) \rangle =\langle s^2 (\tau) \rangle$ and $\langle s^2 ({\bf y}, \tau') \rangle =\langle s^2 ( \tau') \rangle$, which follow from statistical homogeneity.

From Eq.~\eqref{powerfull}, it is straightforward to deduce the behaviour of the function $P(k, \tau, \tau')$ in the IR limit $k \rightarrow 0$. For this purpose, one observes that the r.h.s. of Eq.~\eqref{powerfull} remains finite, because the integral there is converging at spatial infinity $|{\bf x}-{\bf y}| \rightarrow \infty$ by causality. Indeed, correlations of random variables $s({\bf x})$ and $s({\bf y})$ must be 
zero, if they are placed in causally disconnected patches. At equal times, it implies that the separation $|{\bf x}-{\bf y}|$ exceeds the horizon. This should also hold true in the limit $k \rightarrow 0$. Hence, 
\begin{equation}
\label{kindependent}
P(k, \tau, \tau') \left. \right|_{k \rightarrow 0} \rightarrow K(\tau, \tau') \; ,
\end{equation}
where $K(\tau, \tau')<\infty$ is a finite function of times $\tau$ and $\tau'$. This will be important when discussing 
properties of the IR tail of GW spectrum.

The non-linear differential equations may exhibit attractor solutions, so that the asymptotic solution does not depend on the initial conditions, at least on large scales. Moreover, these attractor solutions may reveal self-similarity through dynamic scaling. Assuming such a case for the domain wall network, we restore the time and space dependence of the correlation function on dimensional grounds as 
\begin{equation}
\label{scalingstrict}
\frac{\langle s ({\bf x}, \tau) s ({\bf y}, \tau) \rangle}{\langle s^2 (\tau) \rangle} =F \left(\frac{|{\bf x}-{\bf y}|}{\tau  } \right) \; .
\end{equation}
Any dependence on 
the model dimensionful parameters, that is the domain wall width, is irrelevant once we are interested in the long range dynamics of the domain wall network. 
Namely, the domain wall width plays the similar role as the UV cutoff scale in effective field theories.

Numerical simulations confirm that the domain wall network settles to the scaling regime some 
time after its formation~\cite{Press:1989yh, Hiramatsu:2013qaa}.
One manifestation of the scaling regime is that the number of domain walls in the horizon volume is fixed. Assuming that there is one (a few) domain wall(s) stretching throughout the horizon, we can estimate the energy density of the network as 
\begin{equation}
\rho_{dw} \sim (\sigma_{dw} \cdot  H^{-2} ) /H^{-3} \sim \sigma_{dw} H \; ,
\end{equation}
where $\sigma_{dw}$ is the domain wall tension defined as
\begin{equation}
\sigma_{dw} =\frac{2\sqrt{2\lambda} \eta^3}{3} \; .
\end{equation}
The domain wall energy density is proportional to the Hubble parameter, while the energy density of any dominant component obeying the Friedmann equation is proportional to the squared Hubble parameter. Hence, if the Hubble parameter decreases with time, domain walls eventually dominate the energy density of the Universe, 
unless the tension is chosen to be very small or there 
is a mechanism of domain wall annihilation.

Furthermore, the assumption of the scaling regime imposes important constraints on the form of the correlation function $P(k,\tau, \tau')$ at equal times, which we write as
\begin{equation}
\label{strong}
P(k, \tau, \tau)=\tau^3 \cdot {\cal P} (k\tau) \; ,
\end{equation}
where we introduced the dimensionless reduced power spectrum ${\cal P} 
(k\tau)$. The expression~\eqref{strong} follows from Eq.~\eqref{powerfull}, where we set $\tau=\tau'$ and assume the form of the 2-point correlation function $F$ as in Eq.~\eqref{scalingstrict}. The reduced power spectrum ${\cal P} (k\tau)$ is related to the function $F$ by
\begin{equation}
{\cal P} (k\tau)=\frac{1}{(2\pi)^3} \int d{\bf u} e^{-i{\bf u} \cdot {\bf k} \tau} F(u) \; ,
\end{equation}
where ${\bf u} \equiv {\bf z}/\tau$. Note that the expression~\eqref{strong} could be written on dimensionful grounds and from the scaling considerations that there is the unique mass parameter defining the large scale evolution of the domain wall network --- Hubble rate, or the inverse horizon size. For the same reasons and recalling the symmetry $\tau\leftrightarrow \tau'$, we can immediately write the generic expression for the unequal time correlation function:
\begin{equation}
\label{modest}
P(k, \tau, \tau')=(\tau \tau')^{3/2} \cdot {\cal P} (k\tau, k\tau') \; .
\end{equation}
This consideration has important implications for GWs generated by domain walls, as we show in due course.

\subsection{Energy density of gravitational waves in the scaling regime}

\label{subsec:gwscaling}

GWs sourced by the domain wall network in the scaling regime have a constant energy density. Previously, this conclusion has been obtained from rough estimates based on the Einstein quadrupole formula~\cite{Hiramatsu:2013qaa}. Here we provide a more rigorous analytical derivation of the GW energy density. In what follows, it is assumed that the generation of GWs takes place at radiation domination. 

Equation of motion describing production of GWs is given by 
\begin{equation}
\label{eom}
\left(\frac{\partial^2}{\partial \tau^2} +2 \frac{a'}{a} \frac{\partial}{\partial \tau} -\frac{\partial^2}{\partial {\bf x}^2} \right) h_{ij}=16\pi G a^2 (\tau)\Pi_{ij} ({\bf x}, \tau) \; , 
\end{equation}
(the metric signature is $(+---)$) where $G$ is the Newton's constant, $h_{ij}$ is the transverse traceless (TT) part of the metric perturbation, and 
\begin{equation}
\label{Piij}
\Pi_{ij} ({\bf x}, \tau)=\frac{1}{a^2 (\tau)}\Lambda_{ij, kl} T_{kl} ({\bf x}, \tau) 
\end{equation}
is the TT part of the stress energy tensor; $\Lambda_{ij, kl}$ is the so called Lambda tensor projecting any symmetric tensor into TT form. 

We again switch to Fourier components and perform splitting into polarisations: 
\begin{equation}
\label{split}
h_{ij} ({\bf k}, \tau)=\sum_A e^{A}_{ij} ({\bf k}) h_A ({\bf k}, \tau) \; ,
\end{equation}
and 
\begin{equation}
\Pi_{ij} ({\bf k}, \tau)=\sum_A e^{A}_{ij} ({\bf k}) \Pi_A ({\bf k}, \tau) \; ,
\end{equation}
where the pair of polarisation tensors $e^{A}_{ij} ({\bf k})$ obeys the normalisation condition
\begin{equation}
\label{normalization}
e^{A}_{ij} ({\bf k}) e^{A'}_{ij} ({\bf k})=2\delta_{AA'} \; .
\end{equation}
Initial conditions for Eq.~\eqref{eom} are set at the time $\tau_i$, when 
the domain wall network is formed: $h_{ij}=0$ and $h'_{ij}=0$. With these initial conditions, solution of Eq.~\eqref{eom} in the radiation-dominated Universe reads 
\begin{equation}
\label{sol}
h_{A} ({\bf k}, \tau)=\frac{16\pi G}{a(\tau) k} \int^{\tau}_{\tau_i} 
d\tau' a^3 (\tau') \sin k(\tau-\tau') \Pi_{A} ({\bf k}, \tau') \; .
\end{equation}
The energy density of GWs is given by 
\begin{equation}
\label{GWenergygen}
\rho_{gw} ({\bf x}, \tau)=\frac{1}{32\pi G a^2(\tau)} \langle \frac{\partial h_{ij} ({\bf x}, \tau)}{\partial \tau} \frac{\partial h_{ij} ({\bf x}, \tau)}{\partial \tau}\rangle \; ,
\end{equation}
where averaging is with respect to multiple realisations of a source. 
In Eq.~\eqref{GWenergygen}, we expand the perturbation $h_{ij} ({\bf x}, \tau)$ into Fourier modes $h_{ij} ({\bf k}, \tau)$, then use Eq.~\eqref{split} and substitute the solution~\eqref{sol}. We get 
\begin{align}
\label{superlong}
 & \rho_{gw}(\tau)=\frac{8\pi G}{a^{4}(\tau)}\int d{\bf k}d{\bf q}e^{i({\bf k}+{\bf q}){\bf x}}e_{ij}^{A}({\bf k})e_{ij}^{A'}({\bf q}) \times \\ \notag &\times \int_{\tau_{i}}^{\tau}\int_{\tau_{i}}^{\tau}d\tau'd\tau''a^{3}(\tau')a^{3}(\tau'')\langle\Pi_{A}({\bf k},\tau')\Pi_{A'}({\bf q},\tau'')\rangle\times\\ \notag
 & \times\Bigl[\frac{\sin k(\tau-\tau')\sin q(\tau-\tau'')}{k^{2}\tau^{2}}-\frac{\sin(k\tau+q\tau-k\tau'-q\tau'')}{k\tau}+\cos k(\tau-\tau')\cos q(\tau-\tau'')\Bigr]\,.
\end{align}
From the discussion in Sec.~\ref{subsec:scaling}, where we identify $s ({\bf k}, \tau) \equiv \Pi_A ({\bf k}, \tau)$, one can write for the correlation function on the r.h.s.: 
\begin{equation}
\label{cor}
\langle \Pi_{A} ({\bf k}, \tau') \Pi_{A'} ({\bf q}, \tau'') \rangle =\delta_{AA'} \delta ({\bf k}+{\bf q})  \sqrt{\langle \Pi^2_{ij} (\tau') \rangle}  \sqrt{\langle \Pi^2_{ij} (\tau'') \rangle} \cdot P (k,\tau', \tau'') \; ,
\end{equation}
where we used Eqs.~\eqref{dva} and~\eqref{long}. 

From Eqs.~\eqref{superlong} and~\eqref{cor}, the spectral shape of GWs is given by
\begin{equation}
\label{spectrumanalytical}
\begin{split}
&\frac{d\rho_{gw}}{d\ln k} =\frac{128 \pi^2 G k^3}{a^4 (\tau)} \cdot \int^{\tau}_{\tau_i} \int^{\tau}_{\tau_i} d\tau' d\tau'' a^3 (\tau') a^3 (\tau'') \sqrt{\langle \Pi^2_{ij} (\tau') \rangle}  \sqrt{\langle \Pi^2_{ij} (\tau'') \rangle} \cdot P(k, \tau', \tau'') \times \\ 
& \times \Bigl[\frac{\sin k(\tau-\tau') \sin  k (\tau-\tau'')}{k^2 \tau^2}  -\frac{\sin k(2\tau-\tau'-\tau'')}{k\tau} +\cos k(\tau-\tau') 
\cos k(\tau-\tau'') \Bigr] \; .
\end{split}
\end{equation}
The oscillating term in the square brackets here is independent of $k$ in the limit $k \rightarrow 0$. The same is true for the function 
$P(k, \tau', \tau'')$, as it has been shown earlier, see Eq.~\eqref{kindependent}. We end up with a well-known fact that for $k\tau \ll 1$, the spectral shape of GWs is fixed to be~\cite{Caprini:2009fx, Cai:2019cdl, Hook:2020phx}
\begin{equation}
\label{IRspect}
\frac{d\rho_{gw}}{d\ln k} \propto k^3 \; .
\end{equation}
Let us stress that this behaviour is independent of the assumption of the scaling regime.

Now let us use our assumption of the scaling regime~\eqref{modest}. Furthermore, we assume that the quantity $\sqrt{\langle \Pi^2_{ij} (\tau)\rangle}$ 
evolves with time as the domain wall energy density $\rho_{dw} (\tau) \sim \sigma_{dw} H (\tau)$, i.e., $\langle \Pi^2_{ij} (\tau) \rangle = C\sigma^2_{dw} H^2 (\tau)$, and $C$ is constant in time. This can be inferred from the fact that 
$\rho_{dw} =T_{\mu \nu}u^{\mu} u^{\nu} \sim T_{ij} v^{i} v^{j}=\frac{1}{a^2} T_{ij} (av^{i}) (a v^{j})$. From $\Pi_{ij} \sim \frac{1}{a^2} T_{ij}$, see Eq.~\eqref{Piij}, and using that domain walls move with relativistic velocities, i.e., $av^{i} \sim 1$, we conclude that $\Pi_{ij} \sim \rho_{dw}$. Using the above assumptions, we obtain from Eq.~\eqref{superlong}:
\begin{equation}
\label{const-GW}
\begin{split}
&\rho_{gw} ({\bf x}, \tau)=128 \pi^2 CG \sigma^2_{dw} \int^{\infty}_0 \frac{d\zeta}{\zeta^5} \int^{\zeta}_{\zeta_i} \int^{\zeta}_{\zeta_i} d\zeta'  d\zeta'' ~(\zeta' \cdot \zeta'')^{5/2} \cdot {\cal P} (\zeta', \zeta'') \times \\ 
&\times \Bigl[\frac{\sin (\zeta-\zeta') \sin (\zeta-\zeta'')}{\zeta^2} -\frac{\sin (2\zeta-\zeta'-\zeta'')}{\zeta} +\cos (\zeta-\zeta') \cos (\zeta-\zeta'') \Bigr] \; ,
\end{split}
\end{equation}
where we defined $\zeta \equiv k\tau$ and, consequently, $\zeta' \equiv k\tau'$ and $\zeta'' \equiv k\tau''$. It is straightforward to see that the integral on the r.h.s. converges. It depends on time only through the integration limits $\zeta_i =\zeta \tau_i/\tau$. Since $\zeta_i$ tends to zero in the limit $\tau \gg \tau_i$, thus one can set $\zeta_i =0$ and the integral in this limit does not depend on time:
\begin{equation}
\rho_{gw} (\tau)=\mbox{const} \qquad \tau \gg \tau_i\; .
\end{equation}
In view of the lattice simulations, let us note that the integration with respect to $\zeta$ in practice does not start from zero, because the comoving momentum $k$ 
is typically bounded from below by the box size, $k \geq k_{low}>0$. Physically, this means that we neglect GWs with wavelengths longer than the box. Since the comoving box size and hence $k_{low}$, are constant, the lower integration limit $\xi_{low} \equiv k_{low} \tau$ is growing with time. This may induce a  slight time dependence, which vanishes as the box size goes to infinity.

 Note that the (approximate) constancy of the GW energy density explicit in the form \eqref{const-GW} holds only as soon as the GW background is supported by the domain wall network. After dissolution of the domain wall network, one naturally recovers the standard radiation-like behaviour $\rho_{gw} (\tau) \propto 1/a^4(\tau)$. The spectrum of GWs in terms of the conformal momentum remains intact in this regime. 

\section{Setting up lattice simulations}

Lattice simulations are performed in the cubic box described by $N^3$ lattice points and a constant comoving volume $V=L^3=\mbox{const}$, where $L$ is the comoving box size. The scalar field $\phi$ evolving in the box is equipped with periodic boundary conditions. This box mimics a portion of the early Universe 
expanding with the scale factor $a$ (hence, its physical volume grows as $V_{ph} \propto a^3$) and initially comprising many causally disconnected Hubble patches.
Since $V_{ph}$ is growing slower than the Hubble volume $\propto H^{-3}$, the latter eventually exceeds the former. This is the situation to be avoided, because domain walls tend to escape from the box in that case (recall that there is one domain wall in the horizon volume in the scaling regime). 
Together with a finite lattice spacing $L/N$, this may pose serious obstacles for the reliable interpretation of our simulations. In Sec.~\ref{subsec:optimal}, we discuss the 
technique that allows us to circumvent these limitations and to optimise the parameters of simulations in order to achieve trustworthy results. Before doing that we set up the system to make it suitable for lattice simulations by switching to dimensionless variables and fixing initial conditions in Secs.~\ref{subsec:redef} and~\ref{subsec:initial}.

\label{sec:prep}

\subsection{Field and coordinate redefinitions}

\label{subsec:redef}

In what follows, we switch to dimensionless variables defined as  
\begin{equation}
\tilde{\tau}=\sqrt{\lambda} \eta \cdot \tau\,, \qquad \tilde{x}_i=\sqrt{\lambda} \eta \cdot x_i\,, \qquad \tilde{k}=\frac{k}{\sqrt{\lambda} \eta} \, , \qquad \tilde{\phi}=\frac{\phi}{\eta} \; .
\end{equation}
In terms of dimensionless conformal time, the cosmological scale factor takes the form: 
\begin{equation}
\label{scaleinnewv}
a(\tilde{\tau})=\frac{\sqrt{\lambda} \eta}{H_i} \cdot \frac{\tilde{\tau}}{\tilde{\tau}^2_i} \; ,
\end{equation}
where $H_i$ is the initial Hubble parameter. 
It is convenient to set the initial scale factor $a(\tilde{\tau}_i)$ to unity in what follows: 
\begin{equation}
\label{scaleunity}
a(\tilde{\tau}_i)=1 \; .
\end{equation}
Now, omitting tilde notations, we can write the equation of motion for the field $\phi$ following from the action~\eqref{Lagrangian}:
\begin{equation}
\phi''+\frac{2a'}{a} \phi'-\partial^2_i \phi +\frac{\lambda \eta^2 \tau^2}{H^2_i \tau^4_i} 
\cdot \phi \cdot (\phi^2-1) =0 \;,
\end{equation}
where prime refers to the derivative with respect to the dimensionless conformal time. We are free to start the simulations at the time, when the Hubble rate is given by
\begin{equation}
\label{initialfix}
H_i =\sqrt{\lambda} \eta \; .
\end{equation}
This corresponds to the choice of the initial condition for the Hubble rate. The latter condition is equivalent to setting the initial time of simulations to unity, 
\begin{equation}
\label{tau1}
\tau_i=1 \; ,
\end{equation}
where we took into account Eqs.~\eqref{scaleinnewv} and~\eqref{scaleunity}. Using Eqs.~\eqref{initialfix} and~\eqref{tau1}, one can simplify the equation of motion for the scalar field: 
\begin{equation}
\phi''+\frac{2}{\tau} \,\phi'-\partial^2_i \phi +\tau^2 
\cdot \phi \cdot (\phi^2-1) =0 \; .
\end{equation}
The condition~\eqref{initialfix} is physically justified: while one has $H \gg \sqrt{\lambda} \eta $, where $\sqrt{\lambda} \eta$ is of the order of the field mass, $\phi$ is almost constant being stuck by the Hubble friction. The field configuration starts to evolve only once the Hubble rate drops 
down to the field $\phi$ mass, that explains our choice  \eqref{initialfix}.

Remarkably, evolution of the system is independent of the coupling constant $\lambda$, and thus $\lambda$ can be chosen arbitrary. The same 
concerns the initial Hubble rate $H_i$ as well as the expectation value 
$\eta$, as soon as the condition~\eqref{initialfix} is fulfilled. In particular, we set $\eta=6 \cdot 10^{16}~\mbox{GeV}$, 
when performing lattice simulations. Such a large value is rather unconventional in the setup, where domain walls are supposed to be formed during radiation domination. This choice is, however, purely technical, since
our results can be easily adjusted to any value of $\eta$.

\subsection{Initial conditions for the scalar field}

\label{subsec:initial}

One of the goals of this work is to study the effect of initial conditions 
on evolution of domain wall network. Generically, one writes
\begin{equation}
\langle \phi ({\bf k}) \phi ({\bf q}) \rangle =A(k) \delta ({\bf k}+{\bf q}) 
\end{equation}
and 
\begin{equation}
\langle \dot{\phi} ({\bf k}) \dot{\phi} ({\bf q}) \rangle =B(k) \delta ({\bf k}+{\bf q}) \; .
\end{equation}
As for the first set, we choose vacuum initial conditions with a cutoff $k_{cut}$ applied at high momenta: 
\begin{equation}
\label{vacuum_cond}
A(k)=\frac{\Theta \left(k-k_{cut} \right)}{16\pi^3 k}\,, \qquad \, B(k)=\frac{k}{16 \pi^3} \cdot \Theta \left(k-k_{cut} \right)\; .
\end{equation}
In what follows, we will mostly assume $k_{cut}=\sqrt{\lambda} \eta$, or simply $k_{cut} =1$ in dimensionless units. Strictly speaking, the initial conditions~\eqref{vacuum_cond} correspond to the massless scalar field in the Minkowski spacetime. We do not assume that these correlators correctly describe quantum vacuum of the scalar field in our case, rather for the sake of comparison we follow Ref.~\cite{Hiramatsu:2013qaa}, which was motivated by Ref.~\cite{Felder:2001kt}.

For comparison, for the second set we choose thermal initial conditions: 
\begin{equation}
\label{thermal_cond}
A(k)=\frac{1}{8\pi^3 \cdot k \left(e^{\frac{k}{T}}-1 \right)}\,, \qquad \, 
B(k)=\frac{k}{8\pi^3 \cdot \left(e^{\frac{k}{T}}-1 \right)} \; ,
\end{equation}
where $T$ is the temperature of $\phi$-particles. We choose $T$ to be equal to the Universe temperature for simplicity. Such initial conditions are somewhat less natural in our setup, because one should equip the field $\phi$ with a thermal mass in this case\footnote{We ignore any non-gravitational interactions of the field $\phi$ with primordial bath, cf Ref.~\cite{Blasi:2022ayo} studying the effects of such interactions.}. This is, however, irrelevant for our purposes, because here we aim to check, if the system really approaches the scaling regime, --- an attractor solution, which should be reached independently 
of the starting point. Hence, we are free to pick arbitrary (but clearly distinct) initial conditions.

\subsection{Optimal parameters for lattice simulations}

\label{subsec:optimal}

There are two major challenges when running simulations in the presence of domain walls: 
i) the wall width $\delta_{w}=\sqrt{2/\lambda} \cdot 1/\eta$ must be kept wider than the lattice spacing; 
ii) one must ensure that the horizon volume does not exceed the simulation box volume. 
Optimally these two conditions should be violated at  about the same time $\tau_f$, and then we stop running the code.

The first of the conditions above can be rewritten as 
\begin{equation}
\label{firstconstant}
\delta_{w}=\sqrt{\frac{2}{\lambda}} \frac{1}{\eta}=\frac{2 L_i}{N} \cdot \frac{a(\tau_f)}{a(\tau_i)} \; ,
\end{equation}
where $L_i$ is the physical box size at the moment $\tau_i$. 
The factor $'2'$ on the r.h.s. is introduced such 
that we are restricting to the cases, when the domain wall width is at least twice larger than the lattice spacing.  
The second condition reads 
\begin{equation}
\label{secondconstant}
H^{-1} (\tau_f) =\frac{L_i}{\alpha} \cdot \frac{a(\tau_f)}{a (\tau_i)} \; .
\end{equation}
Here we introduced the ``optimisation'' parameter $\alpha$, which is assumed to be close to unity. Varying this parameter, we will be able to control the level of contamination of the spectrum due to 
IR and UV obstructions of the problem. Here we explicitly assume that the time $\tau_f$ is the same as in Eqs.~\eqref{firstconstant} and~(\ref{secondconstant}). By virtue of Eq.~\eqref{firstconstant}, one can write Eq.\,\eqref{initialfix} as 
\begin{equation}
\label{interconstant}
H^{-1}_i=\frac{1}{\sqrt{\lambda}\eta}=\sqrt{2} \cdot \frac{L_i}{N} \cdot \frac{a(\tau_f)}{a(\tau_i)} \; .
\end{equation}
Dividing Eq.~\eqref{secondconstant} by Eq.~\eqref{interconstant} and using that the Hubble parameter scales with the conformal time as $H\propto\tau^{-2}$ at the radiation domination, one gets 
\begin{equation}
\label{limit}
\frac{\tau_f}{\tau_i} =\frac{1}{2^{1/4}} \sqrt{\frac{N}{\alpha}} \; .
\end{equation}
Then Eq.~\eqref{interconstant} can be rewritten as 
\begin{equation}
\label{constraint}
H^{-1}_i =\frac{1}{\sqrt{\lambda} \eta}=  \frac{2^{1/4} a(\tau_i) L}{\sqrt{\alpha N}} \; ,
\end{equation}
which is the desired constraint on the set of model parameters. Here we have rewritten $L_i$ through $L_i=a(\tau_i) L$; recall that $L$ is the constant comoving box size. 
 In what follows, for the sake of concreteness we will make the following choices of the parameter $\alpha$: 
\begin{equation}
\label{biruv}
\alpha=2\pi~~~\mbox{(basic)} \; , \qquad \alpha=8\pi~~~\mbox{(IR)} \; , \qquad \alpha =\frac{\pi}{18}~~~\mbox{(UV)} \; .
\end{equation}
The choice $\alpha=2\pi$ is the basic one giving rise to the GW spectrum with the IR and UV parts equally respected. 
On the other hand, with the choice $\alpha =8\pi$, one focuses on the IR part of the spectrum somewhat neglecting the UV part, 
and vice versa in the case $\alpha=\pi/18$.

Note that the condition~\eqref{limit} limits the final time of simulations $\tau_f$. In what follows, when discussing GW spectra, we will primarily refer to the lattice with $N=2048$, in which case one has $\tau_f/\tau_i \simeq 15$ for $\alpha =2\pi$.

\section{Results: evolution of the domain wall network}

\label{sec:resultsdw}

\begin{figure}
    \includegraphics[width=0.5\textwidth]{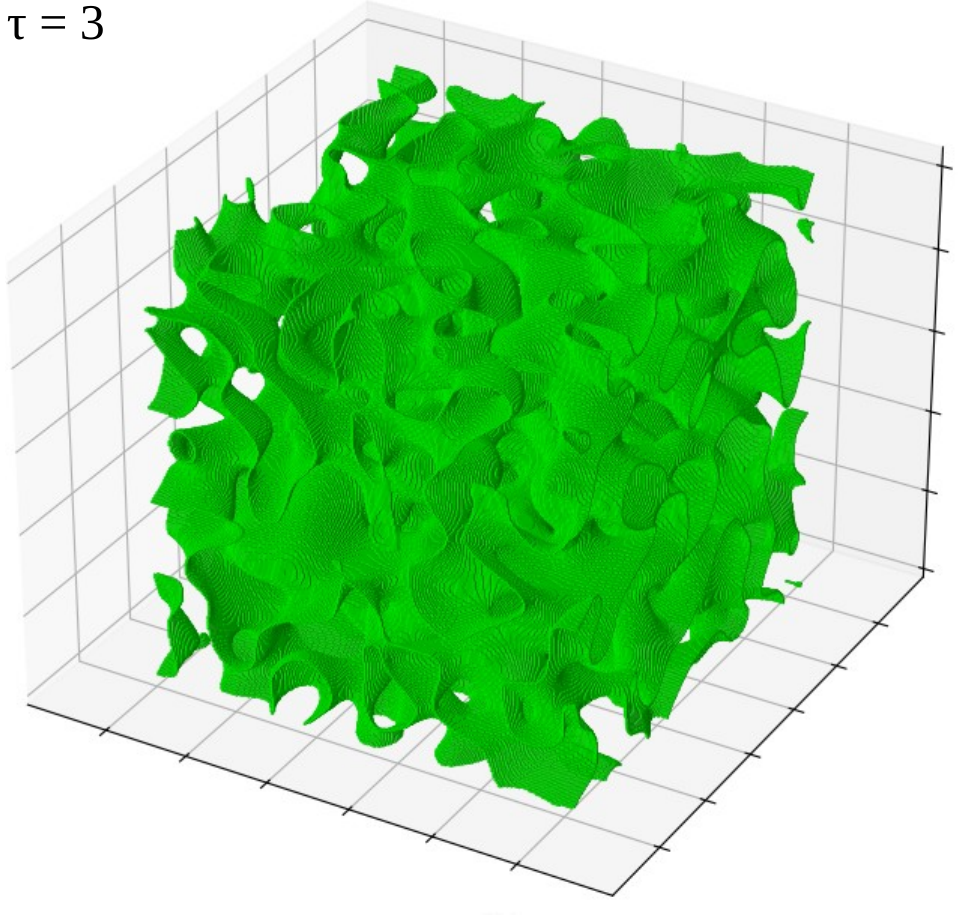}
     \includegraphics[width=0.5\textwidth]{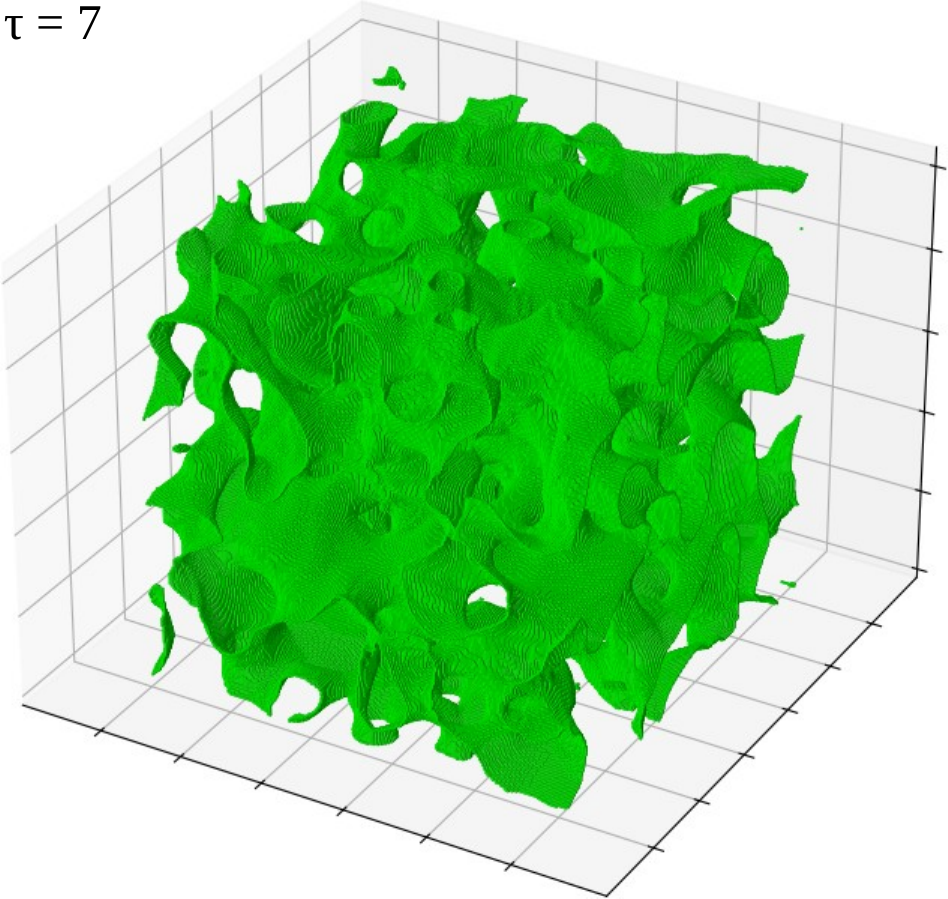}\\
      \includegraphics[width=0.5\textwidth]{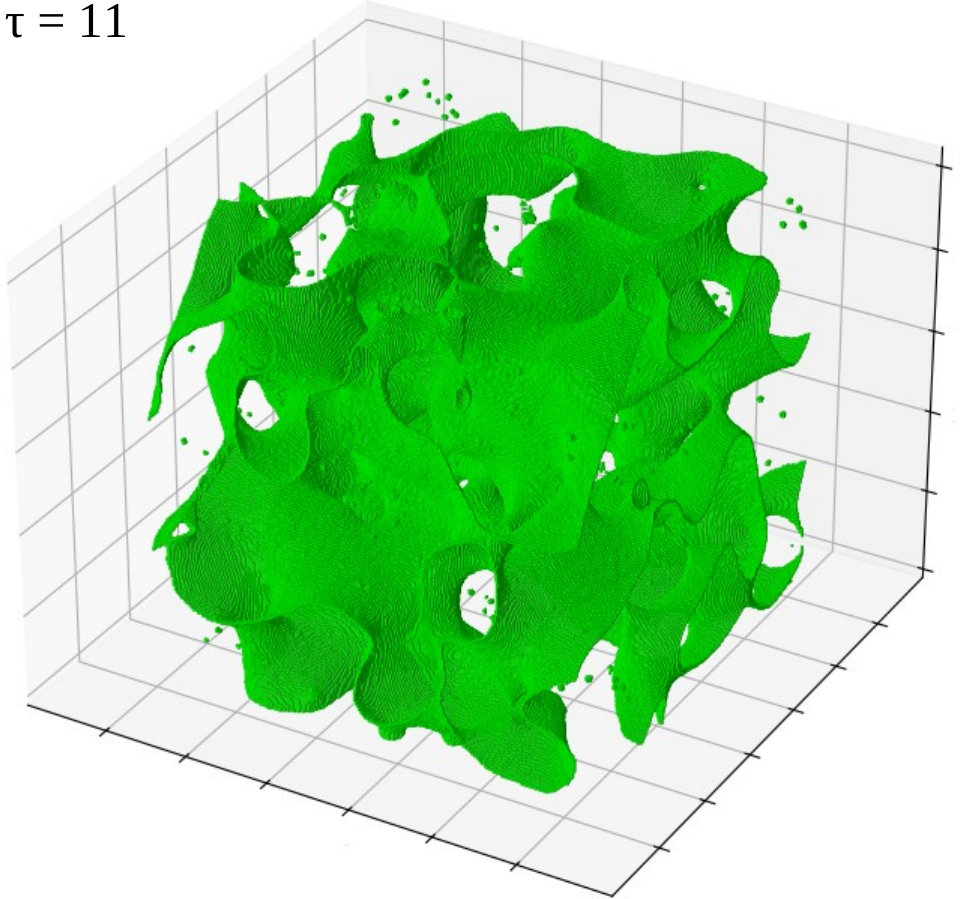}
      \includegraphics[width=0.5\textwidth]{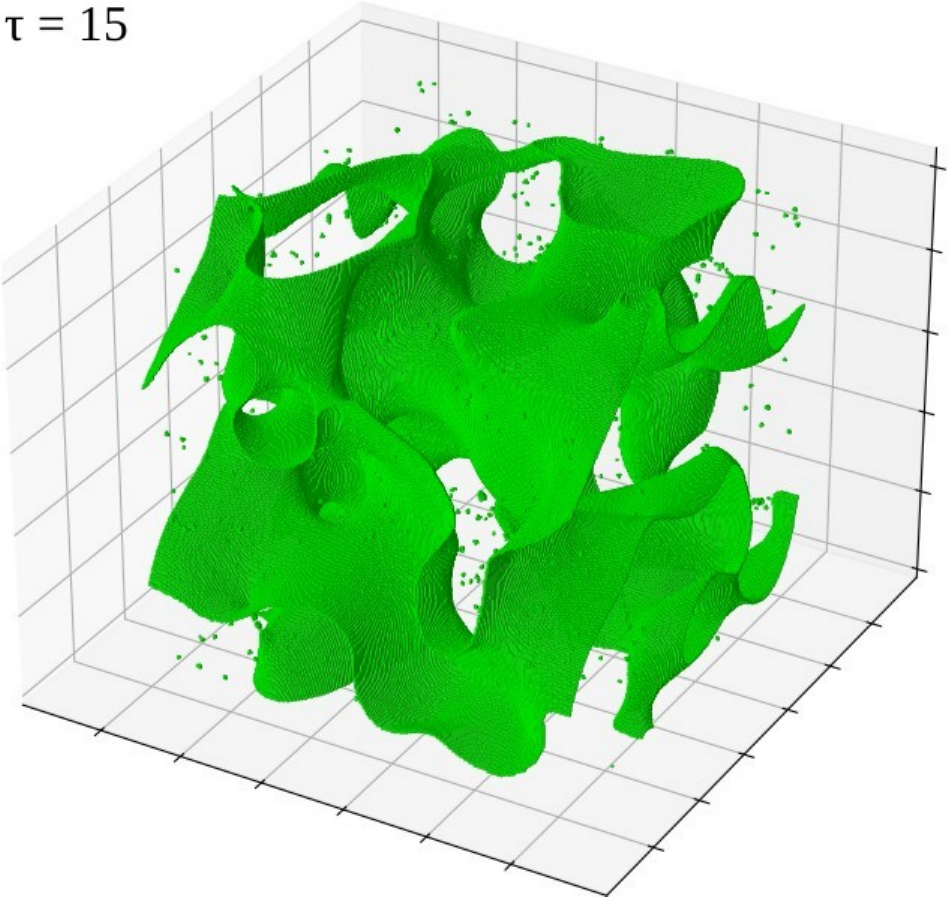}
    \caption{Snapshots of domain wall evolution in the case of vacuum initial conditions at different conformal times $\tau$ in units of $\frac{1}{\sqrt{\lambda}\eta}$. Simulations have been performed starting from vacuum initial conditions on a lattice with the grid number $N=512$. The visible dot-like structures are small size domain walls.}\label{Snapshots_vacuum}
\end{figure}

\begin{figure}
    \includegraphics[width=0.5\textwidth]{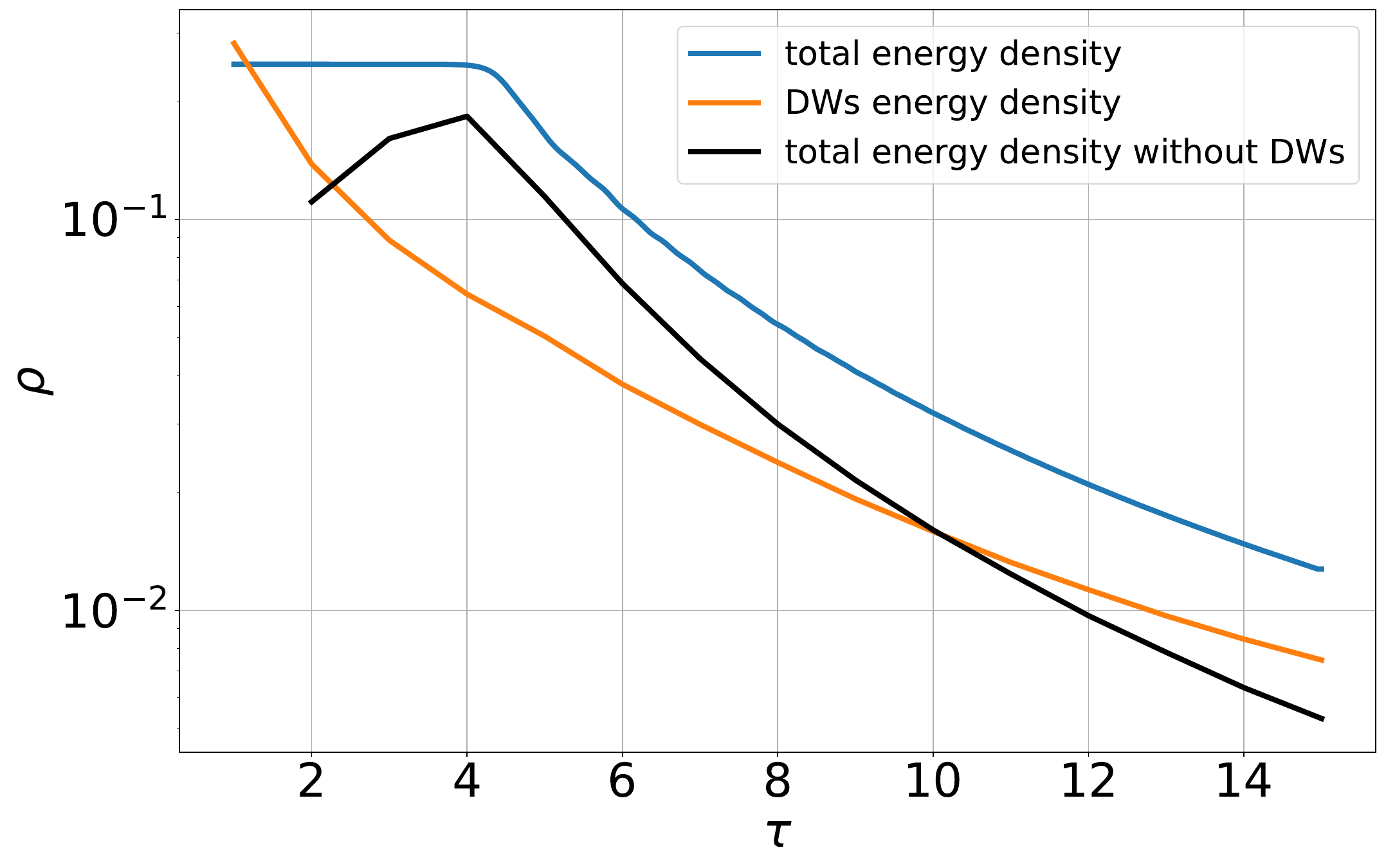}
      \includegraphics[width=0.5\textwidth]{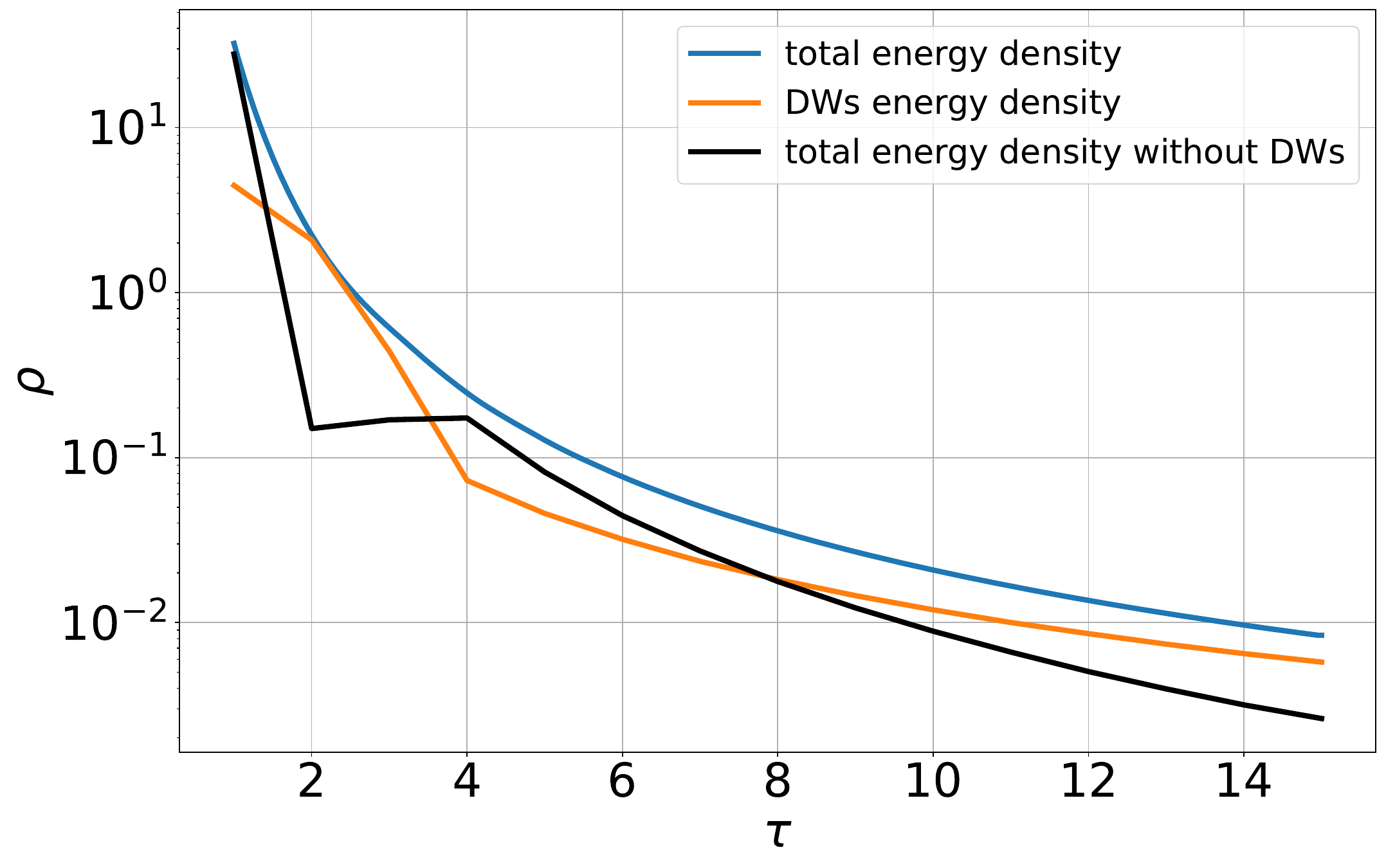}  
    \caption{The energy densities of domain walls and $\phi$-particles are shown with orange and black lines, respectively, in units of $\lambda \eta^4$ for the cases of vacuum (left) and thermal (right) initial conditions. The total energy density of the system is shown with the blue solid line. Conformal time $\tau$ is in units of $\frac{1}{\sqrt{\lambda} \eta}$. Simulations have been performed on a lattice with the grid number $N=512$.} \label{dwenergy}
\end{figure}

Evolution of the domain wall network in the case of vacuum initial conditions is shown in Fig.~\ref{Snapshots_vacuum}. Thermal initial conditions lead to qualitatively similar pictures, and thus they are not shown. One observes that the wall area inside the simulation box decreases, 
and the overall network is mainly comprised of a long single wall and a bunch of small closed walls at late times. We compare the domain wall energy density $\rho_{dw}$ and particles $\phi$ energy density $\rho_{\phi}$ in Fig.~\ref{dwenergy}. Note that evolution of $\rho_{dw}$ can be fitted by $\rho_{dw} \propto 1/a^2(\tau)$, which is in accordance with the estimate of $\rho_{dw}$ in the scaling regime $\rho_{dw} \sim \sigma_{dw} H$. Furthermore, the energy density of $\phi$-particles $
\rho_{\phi} \propto 1/a^3 (\tau)$ seen in Fig.~\ref{dwenergy} is in accordance with Ref.~\cite{Hiramatsu:2012sc} for the case of axion models characterised by the domain wall number\footnote{On the other hand, in the case of axion domain walls the particle energy density $\rho_{\phi}$ decreases slower for larger $N_{dw}$~\cite{Hiramatsu:2012sc}.} $N_{dw}=2$. Such a behaviour of the particles $\phi$ energy density is not quite trivial: it suggests that production of particles $\phi$ is efficient only at early times, while at late times the source of particle production is switched off (see, however, a comment in the end of this section). In particular, this means that closed walls are not efficiently formed during the network evolution, 
so that scalar radiation resulting from their collapse is negligible
(at least in the late time regime). In what follows, we give an additional confirmation 
of this statement. The characteristic time when the energy density of  $\phi$-particles starts tracking the behaviour $\rho_{\phi} \propto 1/a^3(\tau)$ is 
$\tau_{sc} \simeq 4 \tau_i$, which is about the time, when the domain wall network reaches the 
scaling regime, as we show below.

Let us study evolution of the domain wall network more quantitatively. The energy density of the network is given by
\begin{equation}
\label{areap}
\rho_{dw}  \simeq \frac{\sigma_{dw} A }{a(t) V} \; ,
\end{equation}
where $A$ is the comoving domain wall area, and $V$ is the constant comoving volume 
of the simulation box. It is convenient to define the dimensionless scaling parameter $\xi$~\cite{Hiramatsu:2013qaa} also referred to as the area parameter in what follows:
\begin{equation}
\xi \equiv \frac{At}{a(t) V} \; .
\end{equation}
Note that the parameter $\xi$ is constant in the scaling regime. This corresponds to the area $A$ shrinking as $1/\tau$ in the simulation box $V$. We used that $a(\tau) \propto \tau$ and $t =\int a(\tau) d\tau$. Smoothing of the domain wall area seen in Fig.~\ref{Snapshots_vacuum} is a reflection of this fact. 
Conversely, departures of $\xi$ from a constant would signal violation of 
scaling. Numerically, one obtains the area using the formula~\cite{Press:1989yh}: 
\begin{equation}
A=\Delta A \cdot \sum_{\mbox{links}} \delta \cdot \frac{|{\bf \nabla} \phi|}{|\phi_{,x}|+|\phi_{, y}|+|\phi_{, z}|} \; .
\end{equation}
The summation here is over the pairs of grid points (links). The area $\Delta A=(\Delta x)^2$, where $\Delta x$ is the comoving area between the grid points. The quantity $\delta$ takes zero value, if the scalar $\phi$ has same signs at different grid points, and equals unity, if the scalar $\phi$ flips its sign. 

\begin{figure}[t]
    \includegraphics[width=.99\textwidth]{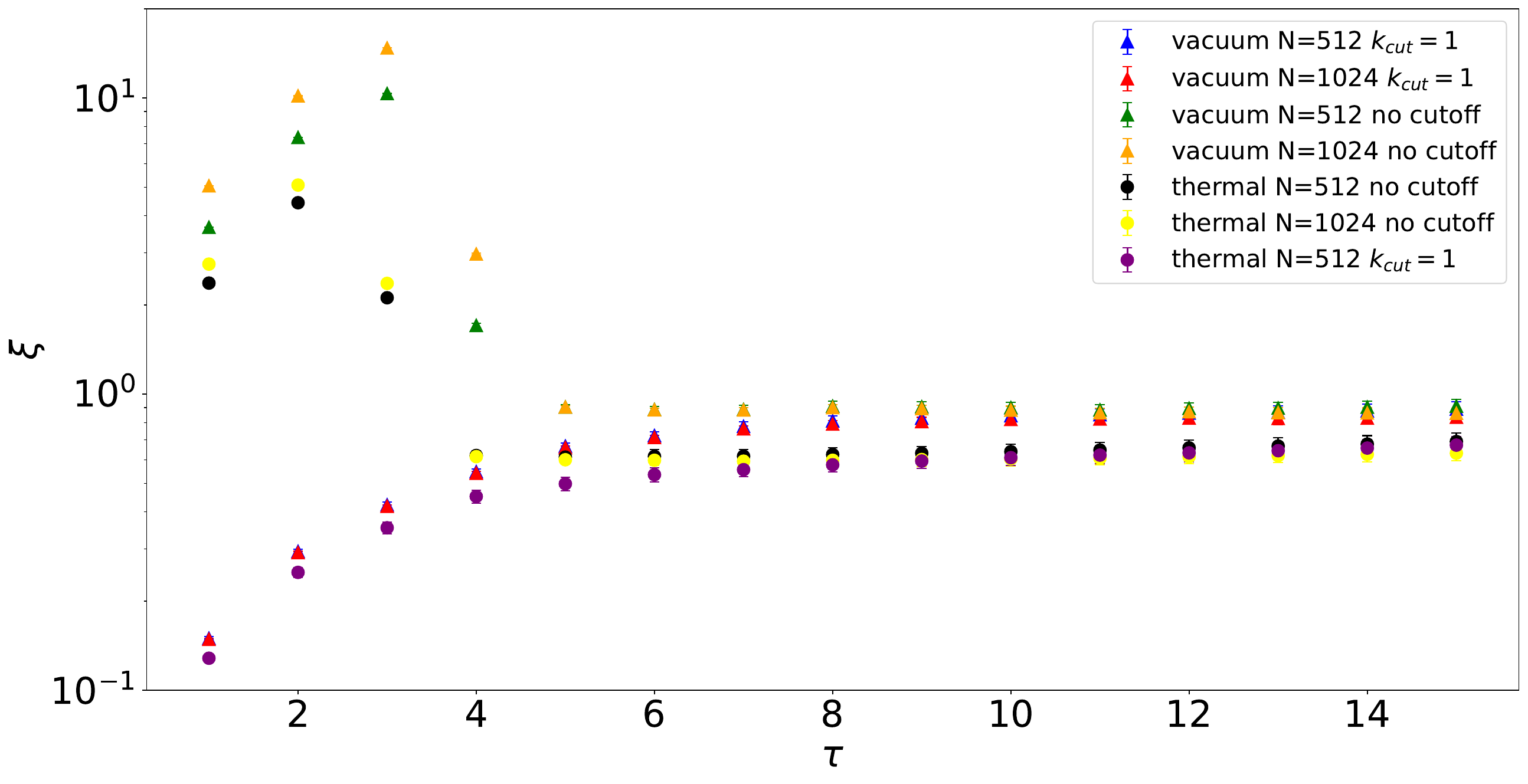}
    \caption{The area parameter $\xi$ inferred in Eq.~\eqref{areap} is obtained from numerical simulations performed on lattices with the grid numbers $N=512$ and $N=1024$ starting from vacuum and thermal initial conditions with and without cutoffs at high momenta. Conformal time $\tau$ and conformal momentum $k$ are in units of $\frac{1}{\sqrt{\lambda} \eta}$ and $\sqrt{\lambda}\eta$, respectively. The parameter $\xi$ taking a constant value reflects that the domain wall 
    network enters the scaling regime. Expectation values and error bars are obtained from 10 simulations run with different base seed values.} \label{ksi}
\end{figure}

We show the results of simulations for the parameter $\xi$ in Fig.~\ref{ksi}. For this purpose we make use of lattices with the grid numbers $N=512$ and $N=1024$ assuming vacuum and thermal initial conditions, 
with and without a cutoff imposed. As the results in cases $N=512$ and $N=1024$ are in a very good agreement between each other, we deem them trustworthy. The area parameter $\xi$ indeed takes a constant value at 
sufficiently late times, and hence the domain wall network enters the scaling regime. Interestingly, the onset of scaling occurs universally at the conformal time $\tau_{sc} \simeq 5$ in dimensionless units, 
or returning to real units, at 
\begin{equation}
\label{scaling_onset}
\tau_{sc} \simeq 5\tau_i \simeq \frac{5}{\sqrt{\lambda} \eta} \; .
\end{equation}
This can be interpreted in terms of domain wall width $\delta_{w}$ relative to the horizon size $H^{-1}$, i.e., the scaling starts, when $\delta_{w}$ constitutes about $\sim 5\%$ of the Hubble radius $H^{-1}$:
\begin{equation}
\delta_{w} \simeq 0.06~H^{-1} (\tau_{sc})\; .
\end{equation}
At the same time, one observes in Fig.~\ref{ksi} that settling to scaling occurs in a drastically different way in the cases with and without a cutoff. While in the former case, the parameter $\xi$ approaches 
a constant from below, in the latter case it approaches from above after a period of initial rapid growth. This can be attributed to the fact that initial scalar fluctuations are larger in the no-cutoff case leading 
to formation of a more curved domain wall initially. 

\begin{figure}[t]
    \includegraphics[width=0.47\textwidth]{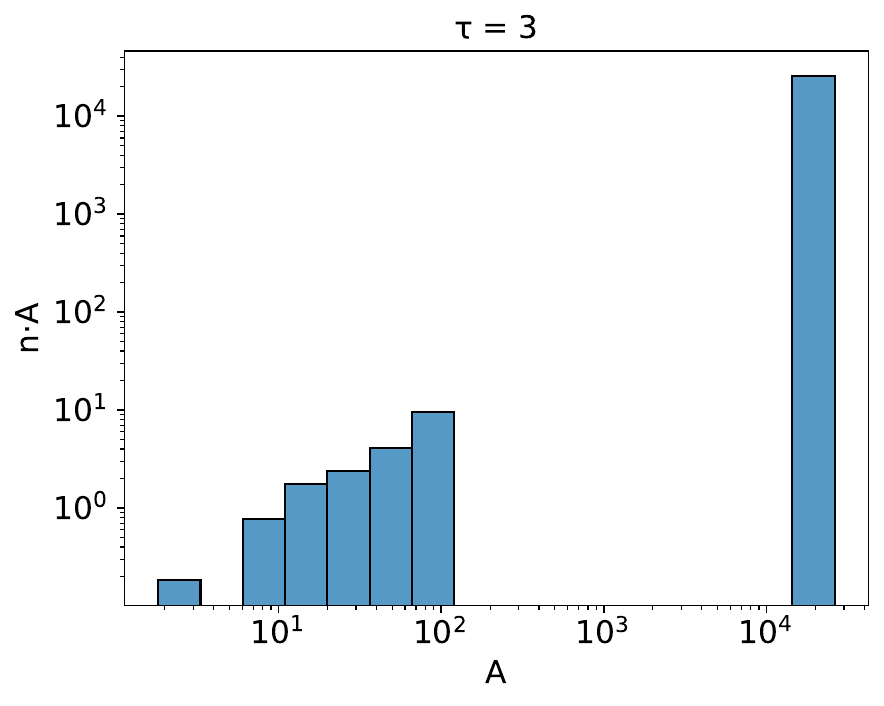}
     \includegraphics[width=0.47\textwidth]{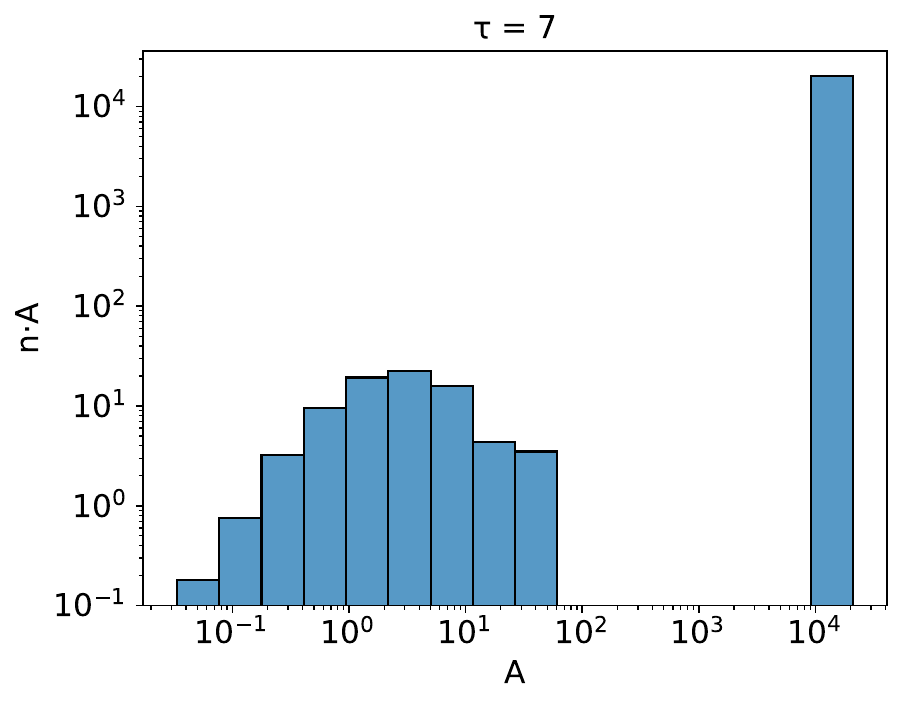}\\
      \includegraphics[width=0.47\textwidth]{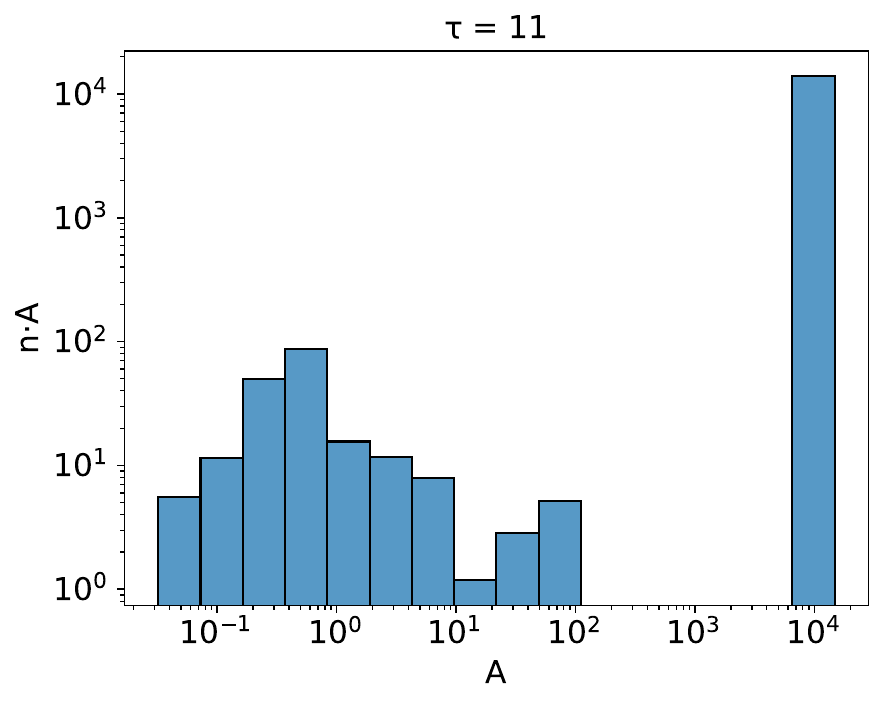}
      \includegraphics[width=0.47\textwidth]{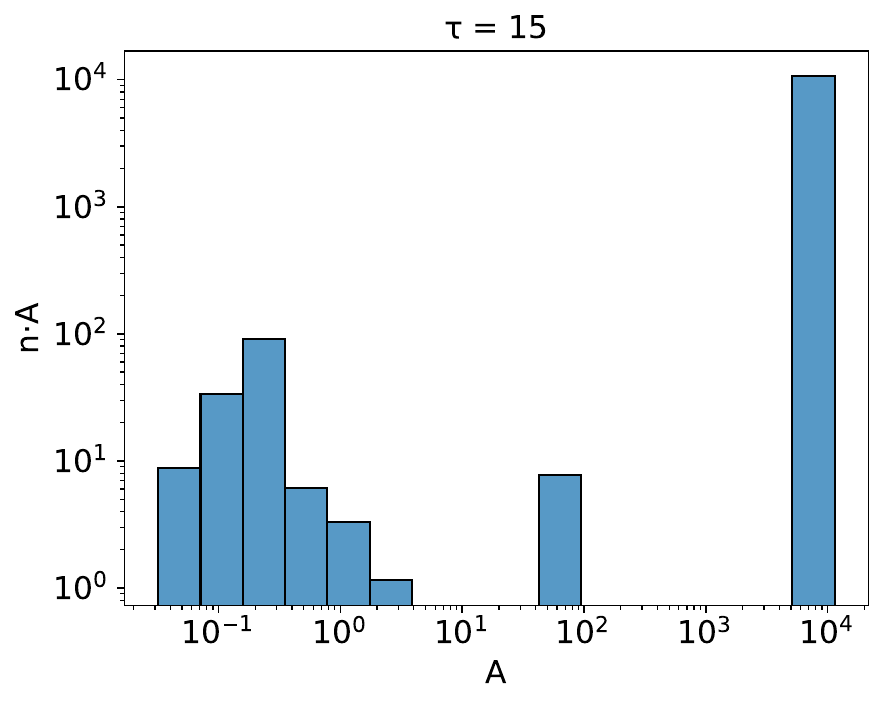}
    \caption{Histograms showing the distribution of the quantity $n \cdot A$ versus $A$, where $A$ is the comoving domain wall area in units of $\frac{1}{\lambda \eta^2}$, and $n$ is the number of domain walls with the area $A$. Distributions are considered at different conformal times $\tau$ in units of $\frac{1}{\sqrt{\lambda}\eta}$. Simulations have been performed starting from vacuum initial conditions on a lattice with the grid number $N=512$. }\label{Histograms}
\end{figure}

Note that the constant values reached by the parameter $\xi$ are clearly distinct in the two cases with thermal and vacuum initial conditions. A similar observation has been made earlier in Ref.~\cite{Kitajima:2023kzu}.
Namely, in the vacuum case, we obtain by fitting to the data
\begin{equation}
\label{ksivac}
\xi = 0.85 \pm 0.04 \qquad (\mbox{vacuum}) \; ,
\end{equation}
while in the thermal case we get 
\begin{equation}
\label{ksither}
\xi = 0.63 \pm 0.04 \qquad (\mbox{thermal}) \; .
\end{equation}
(The error bars correspond to $1\sigma$ CL.) This result is somewhat surprising, because 
the definition of scaling itself implies that the domain wall evolution should be independent of initial conditions. The discrepancy is unlikely to be due to differences in the UV parts of the 
initial scalar spectra. Indeed, analogous spectra with cutoffs imposed lead to the same value of $\xi$, as it is clearly seen in Fig.~\ref{ksi}. We also checked that the discrepancy cannot be 
attributed to the dependence on the finite box size $L$. 
Indeed, the change of $L$ in the case of thermal initial conditions does not lead to considerable changes in $\xi$, leaving it lower than in the case of vacuum initial conditions. This suggests that the discrepancy is likely to have a physical origin.

 As it follows from Fig.~\ref{Histograms}, the total domain wall area is strongly dominated by a single wall stretching throughout the simulation box, while the area of closed walls is negligible. To the best of our knowledge, it is the first time that this statement 
 is clearly demonstrated with histograms. As a consequence, closed domain walls cannot serve as an efficient source of $\phi$-particles. This agrees with the observation from Fig.~\ref{dwenergy} and our discussion in the beginning of this section that the number density of $\phi$-particles is conserved in a comoving volume. At the same time, this appears to be in contrast with the case of cosmic strings, where settling to the scaling regime occurs through formation of loops (analogous to closed walls) evaporating by emission of particles or GWs~\cite{Vachaspati:1984gt, Vilenkin, Hindmarsh:1994re, Blanco-Pillado:2013qja}.

To summarise, we confirm a picture of domain wall network that is essentially comprised of a single long wall when the scaling regime is reached~\cite{Press:1989yh, Hiramatsu:2013qaa}.
In this regime the self-intersection and formation of small closed structures are rare, as it is indicated in Fig.~\ref{Histograms}. The physical area of domain walls captured inside the horizon volume exhibits the growth $\propto H^{-2}$ independently of initial conditions, --- in accordance with the scaling law. At the same time, values of the area parameter shown in Eqs.~\eqref{ksivac} and~\eqref{ksither} are slightly different, suggesting that the scaling solution is sensitive to the choice of initial conditions.

In the end of this section, let us mention one caveat in our discussions. Maintaining the scaling regime generically requires an energy loss of domain walls, which should occur via particle or GW emission. On the other hand, Fig.~\ref{dwenergy} suggests that efficient particle production takes place only before the scaling is reached. Furthermore, GWs discussed in the next section carry a tiny fraction of the network energy density. 
With the current resolution, however, it is difficult to estimate significance of the problem. Indeed, it can be of numerical origin, since the code misses wavelengths longer than the box size or particles with de Broglie wavelength shorter than the lattice spacing.
We postpone a more detailed study of particle emission from domain walls with \CLns~for the future.

\section{Results: gravitational waves}

In this section we take a closer look at GW production as a result of formation and evolution of a network of domain walls discussed above in Sec.~\ref{sec:resultsdw}.
Using the code inbuilt in \CLns~we obtain the total energy density and the spectrum of GWs, depending on initial conditions and parameters of simulations. 
In particular, we check whether the results from analytical approach for the scaling regime presented in Sec.~\ref{subsec:gwscaling} agree with the outcome of numerical simulations.
We identify characteristic properties of different parts of GW spectra and discuss possible new features. 
We also pay a particular attention to distinguishing physical effects in the GW spectra from numerical artefacts associated with the finite size of the grid.

\label{sec:gw}

\subsection{Energy density of gravitational waves}

\label{subsec:efficiency}

\begin{figure}[h]
    \includegraphics[width=\textwidth]{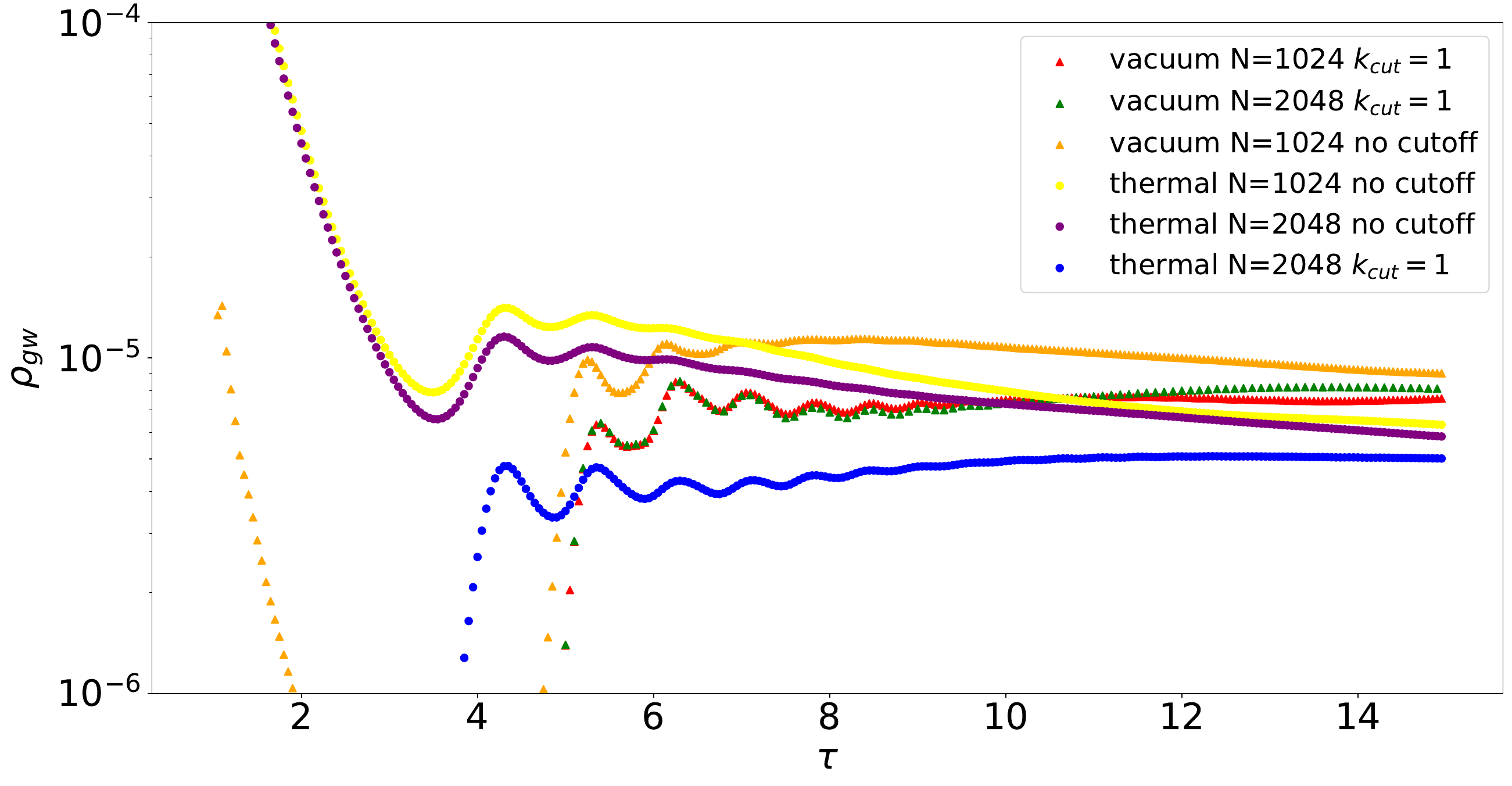}
    \caption{The energy density of GWs in units of $\lambda \eta^4$ emitted by the domain wall network 
    is obtained from numerical simulations on lattices with the grid numbers $N=1024$ and $N=2048$ starting from vacuum and thermal initial conditions with and without cutoffs. Conformal time $\tau$ is in units of $\frac{1}{\sqrt{\lambda} \eta}$. The expectation value $\eta$ is set at $\eta=6 \cdot 10^{16}~\mbox{GeV}$. Rescaling to arbitrary $\eta$ is achieved by multiplying the energy density $\rho_{gw}$ by $(\eta/6 \cdot 10^{16}~\mbox{GeV})^2$.}\label{GWenergydensity}
\end{figure}

We show results for the energy density of GWs $\rho_{gw}$ in Fig.~\ref{GWenergydensity}.
One observes that $\rho_{gw}$ takes a constant value, which reflects the fact that the domain wall network enters scaling regime, in accordance with our discussion in Sec.~2. For the curves corresponding to vacuum initial conditions with the cutoff $k_{cut}=\sqrt{\lambda} \eta$ and for thermal initial conditions, the energy density $\rho_{gw}$ at large $\tau$ takes the values $\rho_{gw} \approx 8.1 \times 10^{-6} \; \lambda \eta^4$ and $\rho_{gw} \approx 5.8 \times 10^{-6} \; \lambda \eta^4$, respectively. 
Making the latter estimates, we primarily 
focused on curves in Fig.~\ref{GWenergydensity} corresponding to $N=2048$. Recalling dependence on model constants and the reduced Planck mass $M_{Pl}=1/\sqrt{8\pi G} \approx 2.44 \cdot 10^{18}~\mbox{GeV}$, i.e., $\rho_{dw} \propto \sigma^2_{dw}/M^2_{Pl}$ (see Eq.~(\ref{const-GW})), and that we have set 
$\eta=6 \times 10^{16}~\mbox{GeV}$, one gets
\begin{equation}
\rho_{gw} \approx \frac{0.015 \cdot \sigma^2_{dw}}{M^2_{Pl}} \; \qquad \mbox{(vacuum)},
\end{equation}
and
\begin{equation}
\rho_{gw} \approx \frac{0.011 \cdot \sigma^2_{dw}}{M^2_{Pl}} \; \qquad \mbox{(thermal)}.
\end{equation}
We see that the GW energy density in the case of thermal initial conditions is slightly lower compared to the case of vacuum initial conditions. We have already observed the similar tendency in Sec.~\ref{sec:resultsdw} for the area parameter $\xi$. 
Furthermore, as in the case of the area parameter, the energy density of GWs approaches a constant from below, if the initial conditions involve a cutoff, and from above, if the cutoff is absent.

\begin{figure}
    \includegraphics[width=\textwidth]{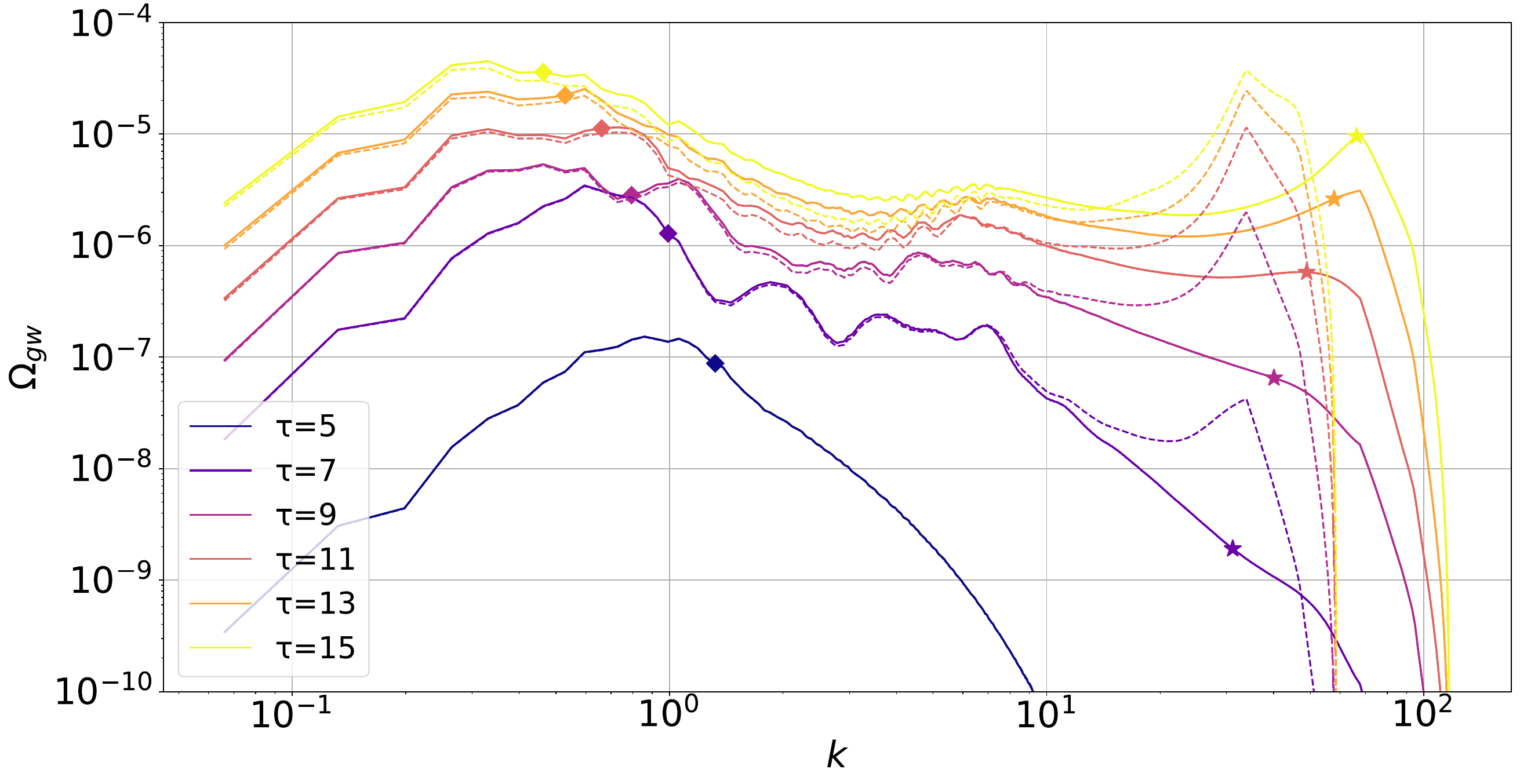}\\
       \includegraphics[width=0.5\textwidth]{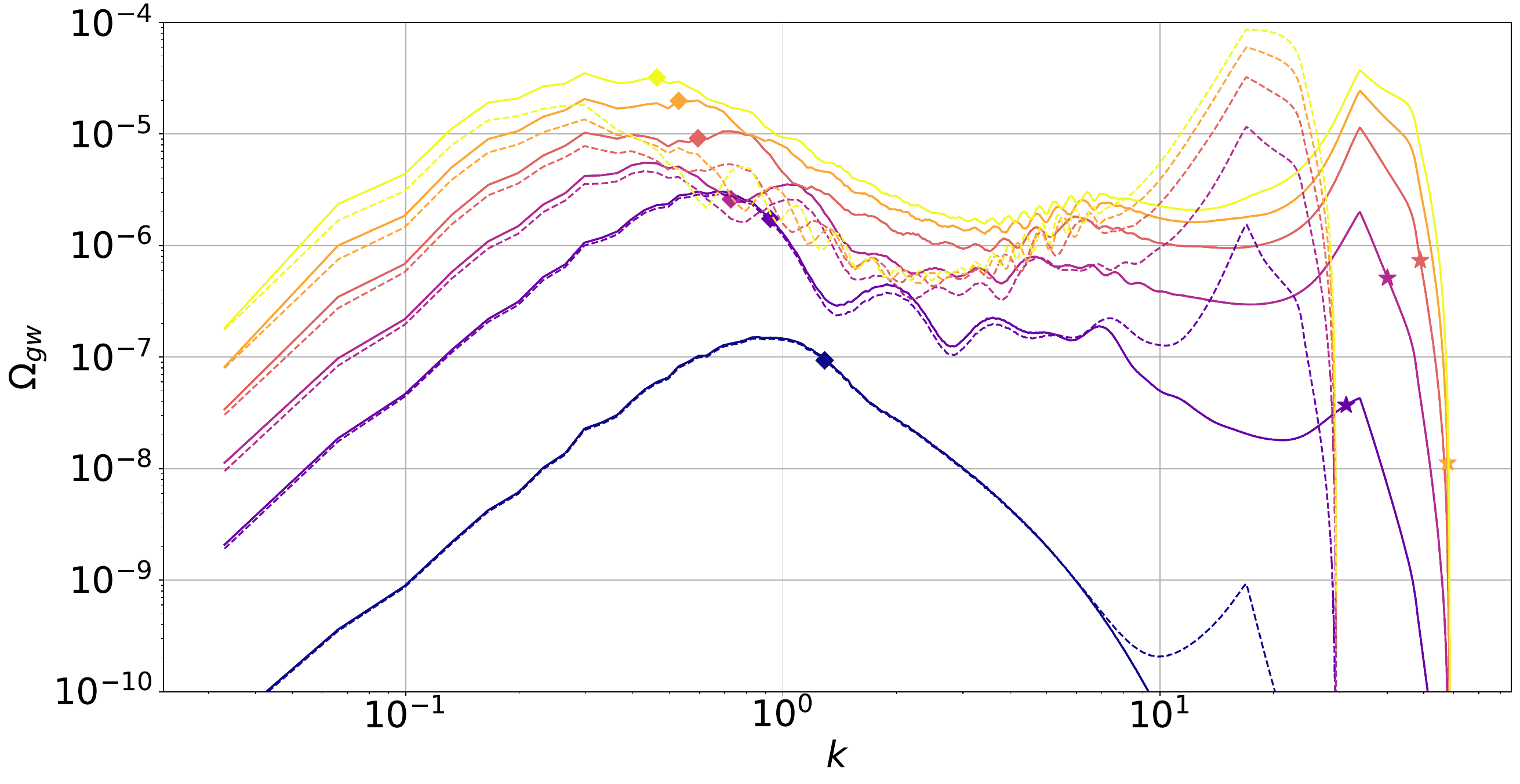}
          \includegraphics[width=0.5\textwidth]{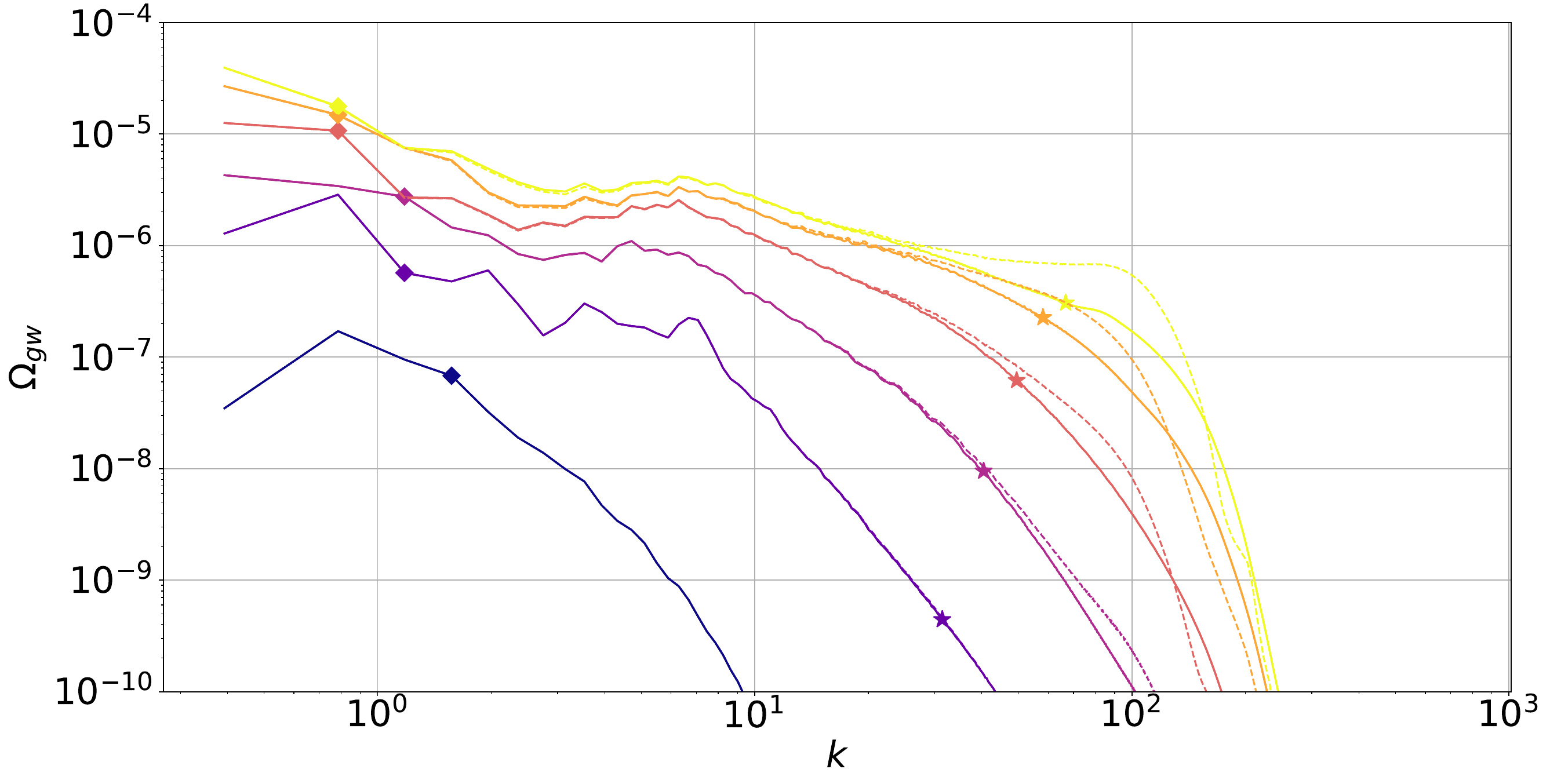}
    \caption{Comparison of GW spectra from the domain wall network obtained by making use of numerical simulations performed on lattices with different grid numbers 
    $N=1024$ (dashed lines) and $N=2048$ (solid lines) starting from vacuum initial conditions with the sharp cutoff set at $k_{cut}=1$. Conformal momenta and conformal times are given in terms of $\sqrt{\lambda} \eta$ and $\frac{1}{\sqrt{\lambda}\eta}$, respectively. The expectation value $\eta$ is set at $\eta=6 \cdot 10^{16}~\mbox{GeV}$. Rescaling to arbitrary $\eta$ is achieved by multiplying the spectra by $(\eta/6 \cdot 10^{16}~\mbox{GeV})^4$. The spectrum shown on the top panel is calculated for the optimal parameters given by Eq.~\eqref{constraint} with the optimisation parameter 
    $\alpha=2\pi$.
    The spectra on the bottom left (right) panel refer to  optimisation parameter $\alpha=8\pi$ $(\alpha=\pi/18)$ allowing us to probe deeper the IR (UV) part of the spectrum. Diamonds correspond to the comoving Hubble scale $k=2\pi H a$, while stars indicate the inverse domain wall width $1/\delta_{w}$, i.e., $k=2\pi a/\delta_{w} $.} \label{GWvacuum}
\end{figure}

\begin{figure}[h]
    \includegraphics[width=\textwidth]{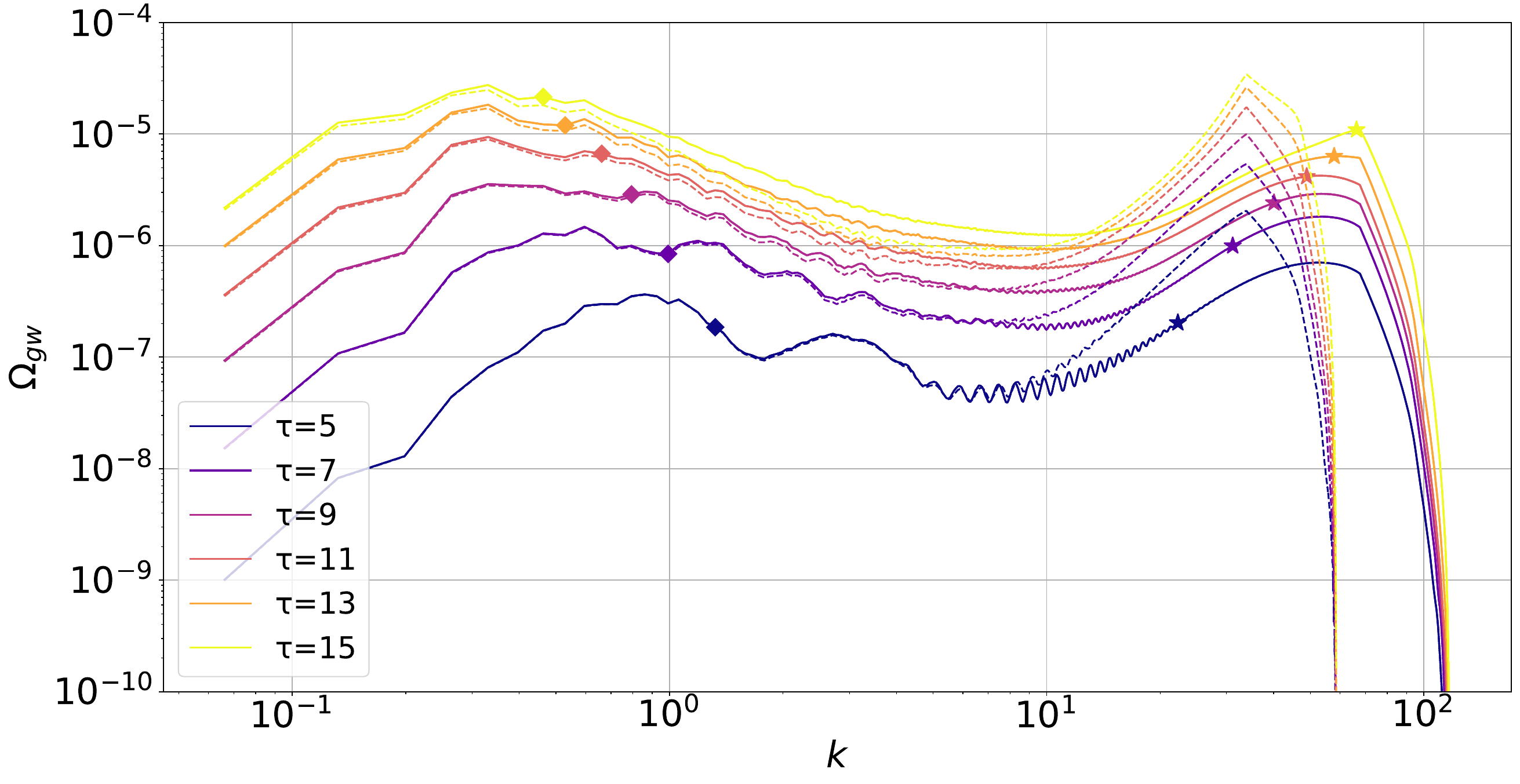}\\
       \includegraphics[width=0.5\textwidth]{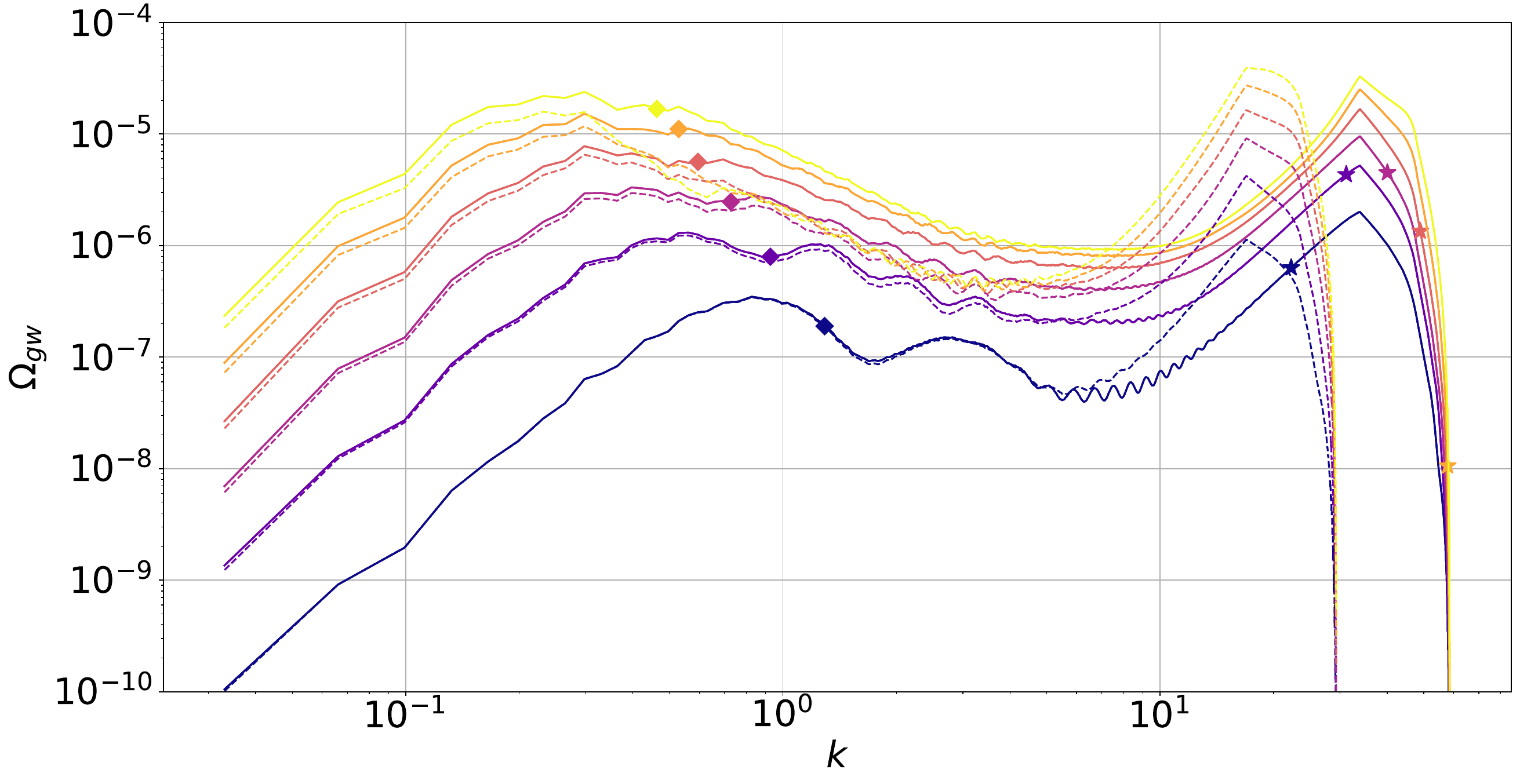}
         \includegraphics[width=0.5\textwidth]{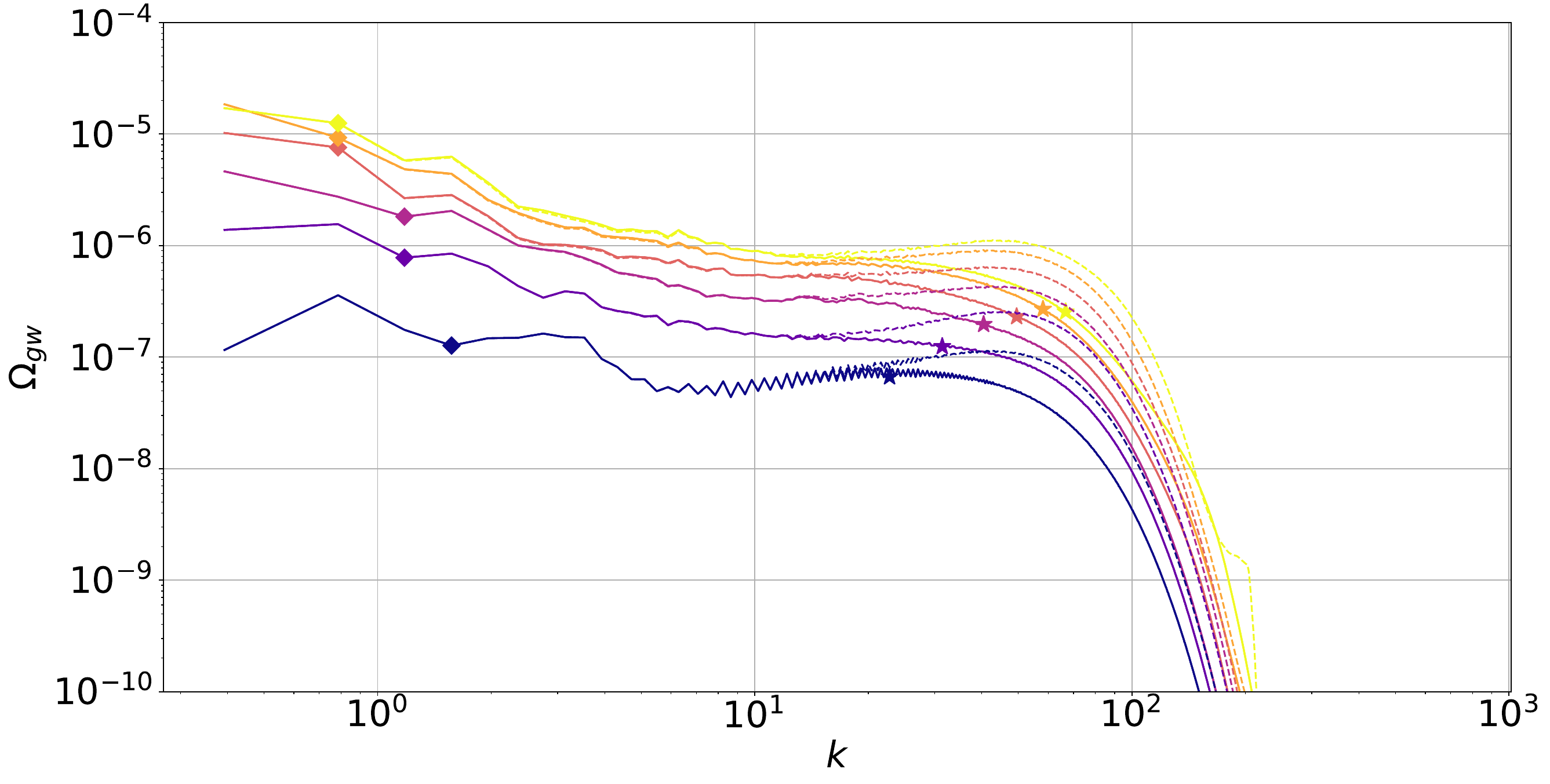}
    \caption{Same as in Fig.~\ref{GWvacuum}, but with thermal initial conditions.}
    \label{GWthermal}
\end{figure}

For comparison with the results of Ref.~\cite{Hiramatsu:2013qaa}, let us quantify the energy density of GWs in terms of the dimensionless quantity: 
\begin{equation}
\label{e}
\epsilon_{gw}=\frac{\rho_{gw}}{G {\xi}^2 \sigma^2_{dw}} \; .
\end{equation}
We infer from our numerical simulations:
\begin{equation}
\label{evac}
\epsilon_{gw} \approx 0.52 \pm 0.05 \;  \qquad \mbox{(vacuum)} \; ,
\end{equation}
and 
\begin{equation}
\label{ethermal}
\epsilon_{gw} \approx 0.70 \pm 0.09 \;  \qquad \mbox{(thermal)} \; .
\end{equation} 
Note that these values are an order of magnitude lower than those in Ref.~\cite{Hiramatsu:2013qaa}, see Fig.~6 there. The origin of this discrepancy is unclear to us, 
taking into account that other results regarding the domain wall evolution, in particular the value~\eqref{ksivac} for the area parameter, and GW spectra discussed below are in a rather good agreement with Ref.~\cite{Hiramatsu:2013qaa}.

\subsection{Gravitational wave spectrum} 

\label{subsec:gw}

Results of numerical simulations of the GW spectrum for different initial  conditions and different lattice resolution are shown in Figs.~\ref{GWvacuum} and~\ref{GWthermal}. Besides the scaling peak at $k\sim 1$ in dimensionless units, one can clearly 
see the peak in the UV part of the spectrum (see top and bottom left panels). Notice, however, that the height and location of the UV peak depend on the lattice resolution: the value of conformal momentum $k$ associated with the UV peak is twice smaller for $N=1024$ as compared to that for $N=2048$. Moreover, the UV peak is absent for the UV-optimised lattice (i.e., when the resolution is increased in the UV by the price of losing predictivity in the IR, see the discussion in Sec.~\ref{subsec:optimal}) shown on the bottom right panels of Figs.~\ref{GWvacuum} and~\ref{GWthermal}. This clearly demonstrates that the UV peak is an artefact of numerical simulations due to a finite lattice spacing and it has no physical origin.

The non-physical UV peak is not localised, but it propagates towards lower frequencies of the GW spectrum due to the non-linear dynamics. This may strongly compromise interpretation of the GW spectrum emerging from numerical simulations. To solve this problem, one observes that the level of contamination depends on lattice resolution. Therefore, parts of the spectrum, which do not change when we modify the lattice resolution, are considered to be trustworthy. Hence, we can rely on the results obtained on lattices with $N=1024$ and $N=2048$, provided only that they overlap. Our method to obtain the GW spectrum is then the following. 
For the vacuum initial conditions we reconstruct central, IR and UV parts of the GW spectrum using different values~\eqref{biruv} 
of the optimisation parameter $\alpha$ entering Eq.~\eqref{constraint}; the results of reconstruction are shown in the top, bottom left, and bottom right panels of Fig.~\ref{GWvacuum}, correspondingly. 
Then we combine these three plots at wavenumbers $k \approx 0.3$ and $k \approx 10$ in dimensionless units to obtain the full GW spectrum, which does not exhibit the pathological peak in the UV.
We repeat the procedure for the thermal initial conditions, combining the central, IR, and UV parts of the GW spectrum shown in the top, bottom left and bottom right panels of Fig.~\ref{GWthermal}, correspondingly, now at wavenumbers $k \approx 0.1$ and $k \approx 7$. Note that such a procedure is legitimate, since the central, IR, and UV parts of the GW spectrum follow from solving the same equation of motion with the same initial conditions in dimensionless units. This is how our main results 
demonstrated in Fig.~\ref{Finalspectra} have been obtained.

From Figs.~\ref{GWvacuum} and~\ref{GWthermal}, we estimate the peak frequency of GWs in both cases of vacuum and thermal initial conditions to be
\begin{equation}
F_{peak} =\frac{k_{peak}}{2\pi a} \simeq 0.7 H \; .
\end{equation}
Namely, the peak frequency flows as the Hubble rate, 
and this flow reflects the domain wall scaling. From the observational point of view, however, only the last instants of GW emission corresponding to the times $\tau \simeq \tau_{dec}$ are relevant. Here the subscript $'dec'$ stands for the moment of time, when the domain wall network collapses. Recall that such a collapse is generically required to avoid overclosing of the Universe. We assume that this moment of time occurs during radiation domination. Taking into account redshifting, the present peak frequency reads: 
\begin{equation}
\label{peakf}
f_{peak} =F_{peak} (\tau_{dec}) \cdot \frac{a_{dec}}{a_0} \simeq 0.7 H_{dec} \cdot \frac{a_{dec}}{a_0} \simeq 7.5~\mbox{nHz} \left(\frac{T_{dec}}{100~\mbox{MeV}} \right) \cdot 
\left(\frac{g_* (T_{dec})}{10} \right)^{1/6} \; ,
\end{equation}
where $a_0$ is the present day scale factor of the Universe. At the last step 
in Eq.~\eqref{peakf}, we have used the normalisation to the decay temperature, 
which is mostly interesting from the viewpoint of PTAs. 
 Interestingly, the inferred peak frequency $f_{peak}$ is slightly smaller than the one obtained in Ref.~\cite{Hiramatsu:2013qaa}, where $F_{peak}$ has been found to be even closer to the Hubble rate. This difference is interesting, but we deem it not to be very significant. 
 
 From Figs.~\ref{GWvacuum} and~\ref{GWthermal} we can also infer the spectral energy density of GWs at the peak: 
\begin{equation}
\label{omegavacuum}
\Omega_{gw, peak} (\tau) \approx 7.7 \times 10^{-10} \cdot \left(\frac{H_i}{H (\tau)} \right)^2 \cdot \left(\frac{\eta}{6 \cdot 10^{16}~\mbox{GeV}} \right)^4 \qquad \mbox{(vacuum)} \; ,
\end{equation}
and
\begin{equation}
\label{omegathermal}
\Omega_{gw, peak} (\tau) \approx 4.6 \times 10^{-10} \cdot \left(\frac{H_i}{H (\tau)} \right)^2 \cdot \left(\frac{\eta}{6 \cdot 10^{16}~\mbox{GeV}} \right)^4 \qquad \mbox{(thermal)} \; .
\end{equation}
These expressions are true during radiation domination prior to the domain wall collapse. Again for the purpose of comparison with Ref.~\cite{Hiramatsu:2013qaa}, we can express $\Omega_{gw, peak}$ as 
\begin{equation}
\label{def}
\Omega_{gw, peak} (\tau) = \frac{8\pi \tilde{\epsilon}_{gw} \xi^2 G^2 \sigma^2_{dw}}{3H^2(\tau)} \; .
\end{equation}
Here following Ref.~\cite{Hiramatsu:2013qaa} we introduce the dimensionless quantity $\tilde{\epsilon}_{gw}$. Comparing Eqs.~\eqref{omegavacuum} and~\eqref{omegathermal} with Eq.~\eqref{def}, one gets 
\begin{equation}
\label{tildeegw}
\tilde{\epsilon}_{gw} \approx 0.25 \pm 0.03 \; .
\end{equation}
We obtain approximately the same value $\tilde{\epsilon}_{gw}$ for both vacuum and thermal initial conditions. Note that the value~\eqref{tildeegw} is in excellent agreement with results of high resolution simulations of Ref.~\cite{Kitajima:2023cek}, see Eq.~(40) there. Furthermore, 
integrating the spectra over $\ln k$, one obtains the energy density $\rho_{gw}$ and the scaling parameter $\epsilon_{gw}$ defined 
in Eq.~\eqref{e} consistent with the values in Sec.~\ref{subsec:efficiency}. 
On the other hand, Ref.~\cite{Ferreira:2023jbu} 
shows higher GW peak power in the case of thermal initial conditions compared to vacuum ones,
contrary to our findings, see Eqs.~\eqref{omegavacuum} and~\eqref{omegathermal}. 
We do not know the source of this contradiction; let us note, however, that the value of $\Omega_{gw, peak}$ in our case appears to be correlated with the domain wall area, or the scaling parameter $\xi$. Namely, larger wall area leads to larger GW signal, consistently with 
naive expectations.

Redshifting the gravitons at matter-dominated and dark energy dominated epochs, one can write for the GW relic abundance: 
\begin{equation}
\Omega_{gw, peak} h^2_0 \approx 1.0 \times 10^{-10} \cdot \left(\frac{100~\mbox{MeV}}{T_{dec}} 
\right)^4\cdot \frac{\sigma^2_{dw}}{(100~\mbox{TeV})^6} \cdot \left(\frac{10}{g_* (T_{dec})} \right)^{4/3} \qquad \mbox{(vacuum)} \; ,
\end{equation}
and 
\begin{equation}
\Omega_{gw, peak} h^2_0 \approx 0.6 \times 10^{-10} \cdot \left(\frac{100~\mbox{MeV}}{T_{dec}} 
\right)^4\cdot \frac{\sigma^2_{dw}}{(100~\mbox{TeV})^6} \cdot \left(\frac{10}{g_* (T_{dec})} \right)^{4/3} \qquad \mbox{(thermal)} \; .
\end{equation}
Here we again adopt the normalisation parameters, which are most suitable from the viewpoint of PTAs. 

\begin{figure}[t]
    \includegraphics[width=\textwidth]{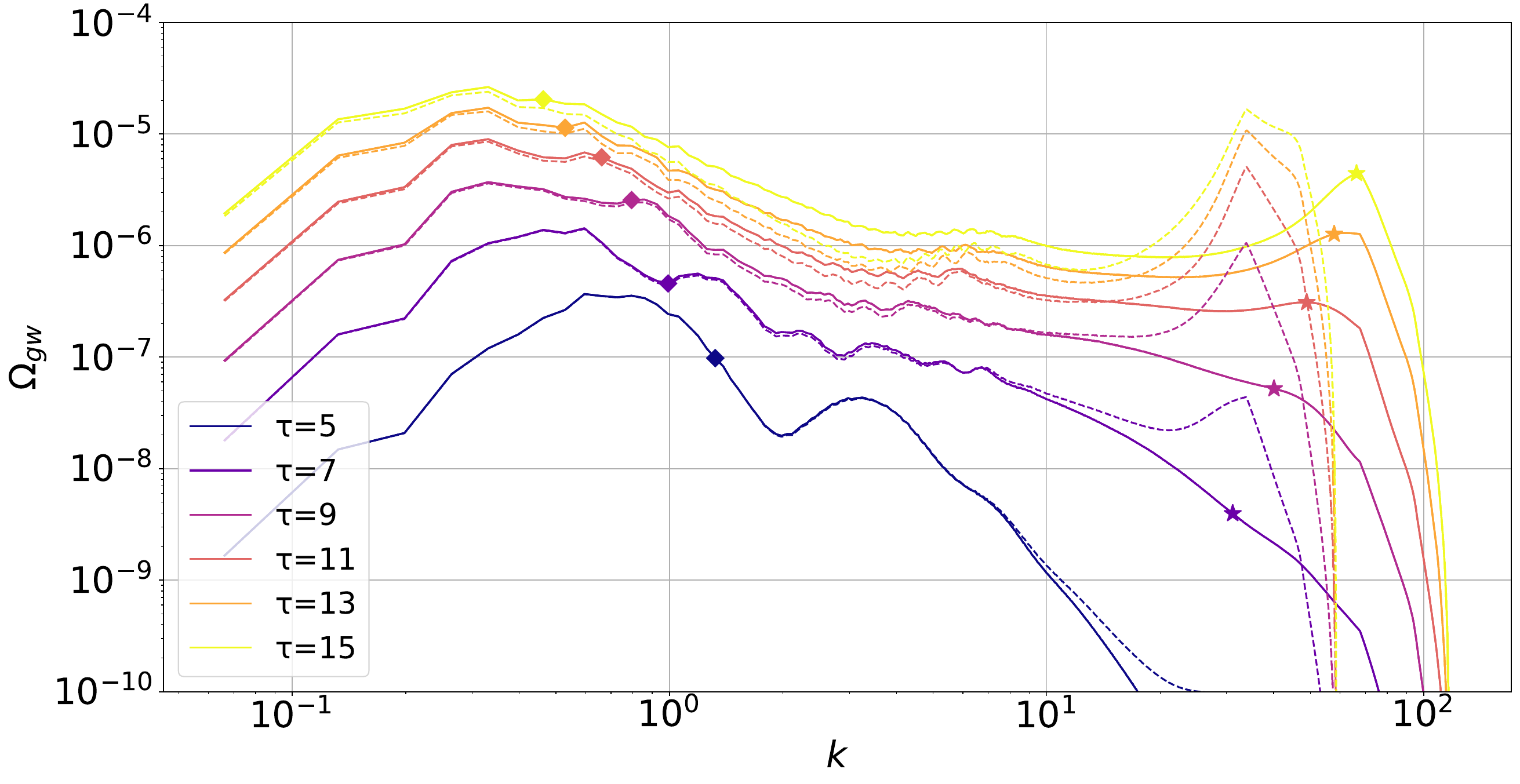}
    \caption{Same as in Fig.~\ref{GWvacuum} (top panel), but for thermal initial conditions with the sharp cutoff set at $k_{cut}=1$.}
    \label{GWthermalcutoff}
\end{figure}

The IR part of the GW spectrum behaves in accordance with causality arguments, i.e., 
\begin{equation}
\Omega_{gw} h^2_0 \propto k^n \qquad k \ll k_{peak} \; ,
\end{equation}
where the exponent $n \simeq 3$, cf. Eq.~(\ref{IRspect}). It is important to stress though that our simulations reveal small, but significant, departures from the exponent $n=3$ suggested by causality. For example, by fitting to the numerical results we obtain $n=2.66 \pm 0.06 $ for vacuum initial conditions and $n=2.69 \pm 0.11$ for thermal initial conditions, respectively, at the moment $\tau = 9$. We attribute this difference to the presence of the low limit for momenta $k$ in the simulations due to a finite box size, while the case $n=3$ is suggested to be recovered in the limit $k \rightarrow 0$. Indeed, 
at the time $\tau=5$, one approaches closer to $n=3$ getting from the fit to simulations $n=2.74 \pm 0.04$ and $n=2.87 \pm 0.06$ for vacuum and thermal initial conditions, correspondingly.

To the right of the peak, the slope of GWs taken at the time $\tau=13$ behaves as (according to the fit to our numerical results)  
\begin{equation}
\label{inter}
\Omega_{gw} h^2_0 \propto k^{-1.53 \pm 0.04}      \qquad k_{peak}<k< k_{*} \qquad \mbox{(vacuum)}\; ,
\end{equation}
and 
\begin{equation}
\label{interm}
\Omega_{gw} h^2_0 \propto k^{-1.28 \pm 0.01}       \qquad k_{peak}<k< k_{*} \qquad \mbox{(thermal)}\; .
\end{equation}
This decrease is sharper than in Refs.~\cite{Hiramatsu:2013qaa, Kitajima:2023cek} giving $-1$ for the exponent, but milder than that of Ref.~\cite{Ferreira:2023jbu} giving $-1.7$. In Eqs.~\eqref{inter} and~\eqref{interm}, we have introduced a new scale $k_{*}$, at which the falloff of the spectrum is followed by a (near) plateau-type of behaviour, i.e., 
\begin{equation}
\Omega_{gw} h^2_0 \simeq \mbox{const} \qquad  k_{*}<k<2\pi a /\delta_{w}\; .
\end{equation}
Note that the plateau region is not perfectly flat, especially in the case of vacuum initial conditions, when there is a bump at intermediate scales. The same observation has been made in Ref.~\cite{Ferreira:2023jbu}.
It appears, however, that the bump is a consequence of the sharp cutoff 
on the vacuum spectrum. Indeed, 
the bump also arises in the case of thermal initial conditions with the cutoff applied, as it follows from Fig.~\ref{GWthermalcutoff}. On the other hand, we see no clear relation between bump locations on the spectra and cutoff scales; therefore the mechanism of a bump creation is not obvious to us. Finally, there is the exponential falloff of the spectrum for the momenta 
exceeding the inverse domain wall width/mass of scalar, i.e., at $k >2\pi a/\delta_{w}$, which has a clear physical origin.

Let us comment on the possible origin of the scale $k_{*}$. As it can be seen in Figs.~\ref{GWvacuum} and~\ref{GWthermal}, the location of $k_*$ approximately stays constant with time, and it is not  sensitive to the choice of initial conditions. That is, the quantity $k_*$ is likely to be related to the Hubble rate and the inverse wall width by 
$k_* \simeq 2\pi \sqrt{H/ \delta_{w}} \cdot a$. This, nevertheless, remains a conjecture, which may be worth investigating in more details in the future. If this hypothesis is correct, then the plateau (bump) is moving deep into the UV region relative to the main  peak at large times. Its width in terms of the ratio of the largest and smallest momenta bounding the plateau is growing with time as $\sqrt{t/\delta_{w}}$. Consequently, the longer the domain wall lifespan is, the better the prospects of observing the plateau in future GW experiments.

\section{Discussion}

\label{sec:discussions}

In this work we revisited evolution of the domain wall network and the generated GW spectrum using the \CLns~code. Since \CLns~is publicly available, 
we believe that the results of this work will pave the way for a broader audience aiming at the analysis of domain walls. The most non-trivial part of such analysis is separating physical effects from artefacts triggered by a limited grid number $N$ and a finite box size. In particular, the artefact caused by finite lattice spacing is manifested as a peak in the UV part of the spectrum. Due to non-linearity of equations, the contamination from the UV peak propagates towards lower momenta with time. Consequently, the longer the time of simulations, the smaller part of the spectrum is available for physical interpretation. We developed a technique allowing to discard nonphysical artefacts and extract trustworthy information from simulations.

After excluding non-physical artefacts we find that our results summarised in Fig.~\ref{Finalspectra} are in a good agreement with those of Refs.~\cite{Kitajima:2023cek, Ferreira:2023jbu}. It is important to mention that unlike Ref.~\cite{Kitajima:2023cek} we did not invoke a potential bias or any other mechanism of domain wall destruction in our simulations. This suggests that the potential bias has little effect on the shape of GW spectrum. However, this statement should be taken with a grain of salt, as some properties of GWs are not yet well established. For example, compared to Refs.~\cite{Hiramatsu:2013qaa, Kitajima:2023cek, Ferreira:2023jbu}, we get slightly different values for the power-law exponent on the r.h.s. of the (scaling) peak, see Eqs.~\eqref{inter} and~\eqref{interm}.
Notably, our values are in between $-1$ and $-1.7$ that were obtained in Refs.~\cite{Hiramatsu:2013qaa, Kitajima:2023cek} and Ref.~\cite{Ferreira:2023jbu}, respectively.

There are some interesting features in the domain wall evolution and GW spectrum revealed after excluding the artefacts, which deserve a closer look. In particular, the area parameter $\xi$ defined in Eq.~\eqref{ksi} takes different values in the cases of thermal and vacuum initial conditions, although it is believed to have same values due to the scaling regime, cf. Ref.~\cite{Kitajima:2023kzu}. We have checked that the difference is unlikely to be attributed to the lattice spacing or the box size. Hence, it may have a physical origin, which is yet to be understood. Furthermore, the spectral energy density of GWs at peak also turns out to be different, which can be related to the aforementioned differences in the parameter $\xi$ evolution. Finally, the origin of the plateau region (bumpy region) at intermediate scales in the GW spectrum is unclear. It appears that the region starts at the comoving scale $k_* \simeq 
2\pi \sqrt{H/\delta_{w}} \cdot a$, but the physical origin of such a scale is lacking.

The presence of the plateau can have interesting consequences for GW searches. In particular, assuming a suitable formation time and a lifespan of the domain wall network, one can have a sizeable GW signal from domain walls in the form of the plateau in the LISA frequency range, which is also predicted 
from cosmic strings and inflationary models. This increases the level of degeneracy between different GW sources, which can be broken, e.g., by GW observations in the nHz frequency range covered by PTAs. Further discussion of potential phenomenological applications of the GW spectra shown in Fig.~\ref{Finalspectra} is out of the scope of this work, and we plan to investigate them in more details elsewhere.

In the future, we are also planning to extend our analysis to include domain walls described by a time-varying tension, with a particular focus on melting domain walls~\cite{Ramazanov:2021eya, Babichev:2021uvl, Babichev:2023pbf}.
The principal advantage of this scenario in comparison to the standard one is that the decreasing domain wall tension solves the problem of overclosing the Universe~\cite{Ramazanov:2021eya, Babichev:2021uvl, Babichev:2023pbf}. Furthermore, the time dependence of the wall tension follows from well motivated particle physics scenarios~\cite{Ramazanov:2021eya}. In this situation, one expects 
a drastically different GW spectrum compared to that of this work. Importantly, the predicted power-law shape of the GW spectrum $\Omega_{gw} \propto k^2$~\cite{Babichev:2021uvl, Babichev:2023pbf, Ramazanov:2023eau} is favoured by the recent PTA observations. It is therefore crucial to perform numerical simulations, similar to those presented in this paper, to support this prediction of GW spectrum in the case of melting domain walls.

\section*{Acknowledgments} 
Numerical simulations were performed on the computer cluster of the Theoretical Division of INR RAS and on the "Phoebe" computer cluster at CEICO/FZU. We are indebted to Josef Dvo\v r\'a\v cek for his assistance in using the latter cluster and to Daniel Figueroa for useful discussions.
I.~D. and D.~G. acknowledge the support of the scientific program of the National  Center for Physics and Mathematics, section 5 "Particle Physics and Cosmology", stage 2023-2025. 
E.~B. acknowledges the support of ANR grant StronG (ANR-22-CE31-0015-01). A.~V. was supported by the Czech Science Foundation (GA\v{C}R), project 24-13079S.

\end{document}